\documentclass{aa}
\usepackage{graphicx}
\begin{document}

\authorrunning{K\"apyl\"a et al.}
\titlerunning{Reynolds stresses and turbulent heat transport}

   \title{Local models of stellar convection:}

   \subtitle{Reynolds stresses and turbulent heat transport}

   \author{P. J. K\"apyl\"a
	  \inst{1}$^{,}$\inst{2},          
          M. J. Korpi
	  \inst{1}$^{,}$\inst{3},
          \and
          I. Tuominen
	  \inst{1}
	  }

   \offprints{P. J. K\"apyl\"a\\
	  \email{Petri.Kapyla@oulu.fi}
	  }

   \institute{Astronomy Division, Department of Physical Sciences,
              University of Oulu, PO BOX 3000, FIN-90014 University of
              Oulu, Finland
	  \and Kiepenheuer-Institut f\"ur Sonnenphysik, 
	      Sch\"oneckstrasse 6, D-79104 Freiburg, Germany
	  \and Laboratoire d'Astrophysique, Observatoire 
	      Midi-Pyr$\acute{\rm e}$n$\acute{\rm e}$es, 
	      14 av. E. Belin, F-31400 Toulouse, France\\ }

   \date{received 16 December 2003 / accepted 14 March 2004}

   \abstract{We study stellar convection using a local
   three-dimensional MHD model, with which we investigate the
   influence of rotation and large-scale magnetic fields on the
   turbulent momentum and heat transport and their role in generating
   large-scale flows in stellar convection zones. The former is
   studied by computing the turbulent velocity correlations, known as
   Reynolds stresses, the latter by calculating the correlation of
   velocity and temperature fluctuations, both as functions of
   rotation and latitude. We find that the horizontal correlation,
   $Q_{\theta \phi}$, capable of generating horizontal differential
   rotation, attains significant values and is mostly negative in the
   southern hemisphere for Coriolis numbers exceeding unity,
   corresponding to equatorward flux of angular momentum. This result
   is also in accordance with solar observations. The radial component
   $Q_{r \phi}$ is negative for slow and intermediate rotation
   indicating inward transport of angular momentum, while for rapid
   rotation, the transport occurs outwards. Parametrisation in terms
   of the mean-field $\Lambda$-effect shows qualitative agreement with
   the turbulence model of Kichatinov \& R\"udiger (\cite{KiRu93}) for
   the horizontal part $H \propto \frac{Q_{\theta \phi}}{\cos \theta}$
   , whereas for the vertical $\Lambda$-effect, $V \propto \frac{Q_{r
   \phi}}{\sin \theta}$, agreement only for intermediate rotation
   exists. The $\Lambda$-coefficients become suppressed in the limit
   of rapid rotation, this rotational quenching being stronger and
   occurring with slower rotation for the $V$ component than for
   $H$. We have also studied the behaviour of the Reynolds stresses
   under the influence of a large-scale azimuthal magnetic field of
   varying strength. We find that the stresses are enhanced by the
   presence of the magnetic field for field strengths up to and above
   the equipartition value, without significant quenching. Concerning
   the turbulent heat transport, our calculations show that the
   transport in the radial direction is most efficient at the
   equatorial regions, obtains a minimum at midlatitudes, and shows a
   slight increase towards the poles. The latitudinal heat transport
   does not show a systematic trend as a function of latitude or
   rotation.

  }

   \maketitle


\section{Introduction}

   The Sun is rotating differentially; its internal rotation profile
   is known from helioseismic inversions (e.g. Schou et
   al. \cite{Schou98}, Thompson et al. \cite{Thompson03}). There also
   exists a meridional large-scale flow, although the magnitude and
   structure of this meridional circulation pattern is rather poorly
   known observationally (e.g. Stix \cite{Stix02}). For more active
   rapidly rotating stars, photometric (e.g. Hall \cite{Hall91}; Henry
   et al. \cite{Henry95}; Donahue et al. \cite{Dona96}) and
   spectroscopic (e.g. Collier Cameron et al. \cite{CoCa02}; Reiners
   \& Schmitt \cite{ReSch03a}, \cite{ReSch03b}) observations have been
   used to deduce the surface differential rotation. These
   observations show that the relative differential rotation $\Delta
   \Omega / \Omega$ strongly decreases as a function of rotation,
   whilst the absolute one, $\Delta \Omega$, remains roughly constant.

   The solar (and to some extent stellar) rotation has been studied
   using three main approaches: local and global convection models and
   mean-field calculations. The global convection models, which have
   been in use for more than two decades (e.g. Gilman \cite{Gilman77})
   improving in the coverage of spatial scales, have only very
   recently started to show reasonable agreement with the helioseismic
   observations (e.g. Robinson \& Chan \cite{RobCha01}; Brun \& Toomre
   \cite{BruTo02}). The clear advantage of these models is the fact
   that the Reynolds stresses, meridional flows, differential rotation
   and their non-linear interactions can be studied from the same
   model. These calculations, however, are computationally expensive,
   making a comprehensive study of different types of stars with
   varying rotation and depths of convection zones difficult to
   undertake. Furthermore, the spatial scales covered are not yet
   adequately small to consistently model the small-scale turbulence,
   which may have an important influence on the dynamics of the
   large-scale flow.

   A theory of the role of anisotropic turbulence on angular momentum
   transport and large-scale motions was formulated by R\"udiger
   (\cite{R77}, see also R\"udiger \cite{Rudiger89}, hereafter R89,
   for a thorough discussion), in the form of the so-called
   $\Lambda$-effect: in a rotating anisotropic fluid additional
   Reynolds stresses arise, which are proportional to $\Omega$ itself,
   therefore non-vanishing even for rigid rotation; in stellar
   convection zones the radial density stratification provides a
   natural source of the required anisotropy. These turbulent
   stresses, as an ensemble average, lead to angular momentum
   transport both horizontally and vertically, giving rise to
   non-uniform rotation profiles. This can be quantified, under the
   mean-field approximation, as the components of the $\Lambda$-tensor
   in the series expansion $Q_{ij}=\Lambda_{ijk} \Omega_k +
   \mathcal{N}_{ijkl} \partial \left< u_k \right> / \partial
   x_l$. However, a major difficulty remains: how to determine the
   magnitude and form of the tensor components for the convectively
   turbulent flow, for which no complete theory exists.

   The small-scale turbulent motions can also take part in the
   generation of meridional flows, due to the rotationally induced
   latitude dependence of the turbulent heat flux, measured by the
   correlation tensor $\langle u'_i T'\rangle$ (e.g. R89). The
   meridional component of the Reynolds stress tensor, $Q_{r \theta}$,
   can also have an influence on the meridional flows (e.g. R89),
   although the significance of this component has not yet been fully
   explored. Any meridional flow generated will take part in the
   angular momentum transport, capable of driving horizontal
   differential rotation; therefore, its investigation is of
   importance in determining the stellar rotation profiles.

   In most cases the parametrisations for the turbulent quantities are
   derived under some simplifying assumptions, most notably the first
   order smoothing approximation (FOSA), also called the second order
   correlation approximation (SOCA) or the quasi-linear approach, in
   which no terms higher than second order in the linearised equations
   for the fluctuating fields are taken into account. This is valid in
   the hydrodynamic case if the Strouhal number, which is the ratio of
   the coherence time to the turnover time, is small. This condition
   is not very well satisfied in the stellar environment (e.g. Stix
   \cite{Stix02}, see also Brandenburg et
   al. \cite{Brand04}). Nevertheless, models applying FOSA have been
   used to derive the $\Lambda$-tensor coefficients (recently e.g. by
   Kichatinov \& R\"udiger \cite{KiRu93}, hereafter KR93), the
   diffusive part of the turbulent stresses (Kitchatinov et
   al. \cite{Kitcha94a}), and to investigate the behaviour of the
   $\Lambda$-coefficients in the magnetised regime (Kitchatinov et
   al. \cite{Kitcha94b}).

   The resulting formulations of the transport coefficients have been
   used to study stellar rotation with mean-field models, in which
   approach, in the simplest form, only the mean-field momentum
   (Reynolds) equation is solved for (e.g. Kitchatinov et
   al. \cite{Kitcha94a}). More sophisticated models (e.g. Brandenburg
   et al. \cite{Brand92}) work with the MHD equations taking into
   account the thermodynamics as well. The advantage of these models
   is the low computational cost, allowing for the exploration of
   large parameter regimes. Using this approach, rotation laws of the
   young solar-like stars and the Sun, as well as some other stars
   (e.g. Kichatinov \& R\"udiger \cite{KiRu93}, \cite{KiRu95},
   \cite{KiRu99}; K\"uker et al. \cite{Kuker93}; K\"uker \& Stix
   \cite{KuSti01}) have been calculated. Along a different line of
   study, Rieutord et al. (\cite{Rieutord94}) demonstrated that the
   differential rotation from a direct convection calculation in
   spherical coordinates agrees with a mean-field model using a
   $\Lambda$-effect derived from direct calculation.

   Another way to derive the turbulent transport coefficients,
   bypassing the difficulties of solving the full problem in a
   spherical domain and the probably oversimplified analytical
   techniques, is to restrict the calculations to a small Cartesian
   domain, located at different latitudes and depths in the stellar
   convection zone. Following this kind of an approach, Hathaway \&
   Somerville (\cite{Hatha83}), Pulkkinen et al. (\cite{Pulkki93}),
   Brummell et al. (\cite{Brumm98}), and, in a more comprehensive
   study, Chan (\cite{Chan01}), have calculated the off-diagonal
   components of the Reynolds stress tensor. In the study of Pulkkinen
   et al. (\cite{Pulkki93}) the turbulent heat transport was
   investigated as well.

   The investigation of solar and stellar rotation plays an important
   role in understanding the stellar magnetic activity as well. This
   is because the solar and stellar magnetic fields are believed to
   arise due to the inductive action of turbulent motions and
   differential rotation, which is generally called the turbulent
   dynamo mechanism. In the special case of the Sun, for which the
   rotation profile is an observed quantity, the dynamo models
   succesfully reproduce the solar magnetic activity patterns only if
   a relatively strong meridional circulation is included
   (e.g. Bonnano et al. \cite{Bon02}); since the observational
   knowledge of such motions is still rather poor, it is important to
   study them from numerical models both on global and local
   scales. For the more active rapidly rotating stars, observational
   information is limited to the deduction of the surface differential
   rotation, giving an even more important role for the numerical
   investigations.

   In this study, we investigate the characteristics and correlations
   in a turbulent convective flow, using a local three-dimensional MHD
   model of Caunt \& Korpi (2001), on top of which we have implemented
   a three-layer setup consisting of an overshoot layer on the bottom,
   a convectively unstable layer in the middle, and a narrow cooling
   layer on top of the domain. As in the previous studies, we
   calculate the off-diagonal components of the Reynolds stresses and
   radial and meridional components of the turbulent heat transport
   tensor, but extend our calculations further to the rapid rotation
   end (up to Coriolis number 15), and also investigate the influence
   of imposed magnetic fields on the turbulent dynamics. As our model
   includes a stable overshoot layer below the convectively unstable
   one, we also report on the relevance of this layer for the
   turbulent dynamics, and in addition on the dependence of the
   overshooting efficiency as a function of rotation and latitude.

   The remainder of the paper is organised as follows: in
   Sect.\,\ref{sec:model} the model is described, in
   Sect.\,\ref{sec:cameth} the parameters of the calculations are
   given and the methods used are discussed, and in
   Sect.\,\ref{sec:theLaTH} the mean-field theory of the
   $\Lambda$-effect and turbulent heat transport is
   introduced. Sects.\,\ref{sec:results} and \ref{sec:conclu} present
   the results and the conclusions, respectively.

\section{The model}
\label{sec:model}

   \begin{figure*}
   \centering
   \includegraphics[width=0.9\textwidth]{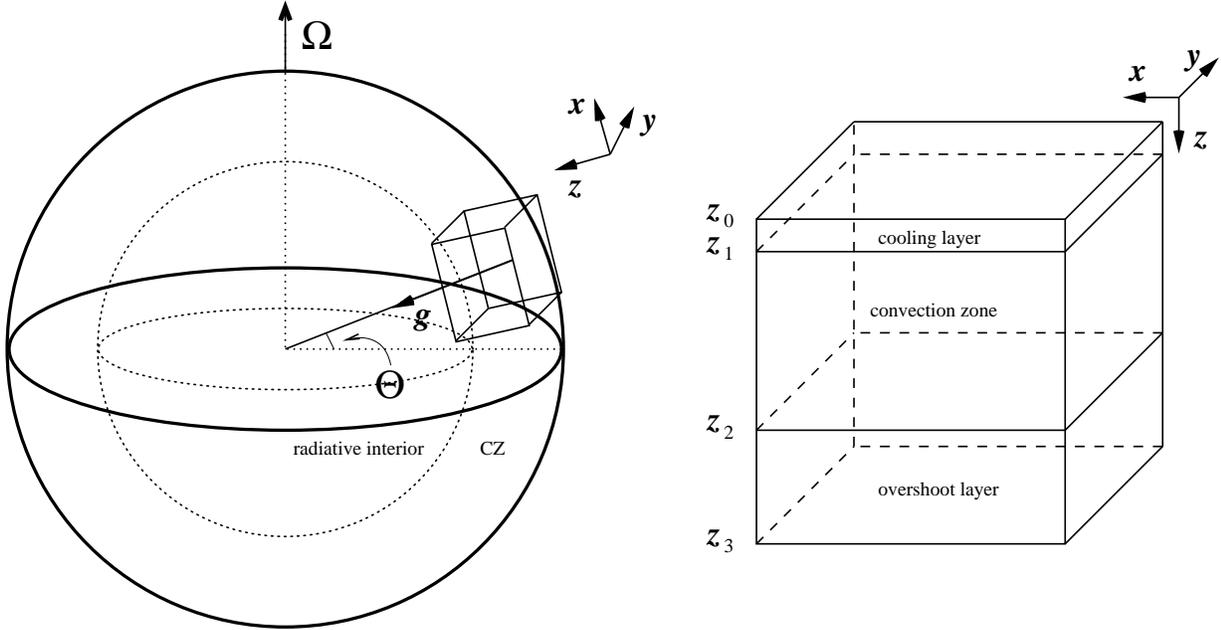}
      \caption{Sketch of the model setup. Left: the coordinate system
      is set up so that $x$-axis points from south to north, $y$-axis from
      west to east, and $z$-axis radially inwards. The angle $\Theta$ is the
      latitude, i.e. the angular distance of the centre of the box
      from the equator. Right: the domain is divided into three parts,
      an upper cooling layer, the convectively unstable layer, and the
      lower overshoot region. The coordinates $z_{0}$ and $z_{3}$
      denote the upper and lower boundaries of the box, whereas
      $z_{1}$ and $z_{2}$ are the boundaries between stable and
      unstable layers.}
      \label{pic:model}
    \end{figure*}

   The computational domain in our calculations is a rectangular box,
   situated on the southern hemisphere of a star at a latitude
   $\Theta$, see Fig.~\ref{pic:model}. The coordinates are chosen so
   that $x$, $y$, and $z$ correspond to the south-north, west-east,
   and radially inward directions, respectively. When analysing the
   results it is important to keep in mind that our $(x,y,z)$
   corresponds to $(-\theta,\phi,-r)$ in spherical polar coordinates, 
   where $\theta$ is the colatitude. The
   angular velocity as function of latitude can now be written as
   $\vec{\Omega} = \Omega (\cos \Theta \vec{\hat{{e}}_{x}} - \sin
   \Theta \vec{\hat{{e}}_{z}})$. We solve a set of MHD-equations
   \begin{eqnarray}
   \frac{\partial {\vec A}}{\partial t} &=& {\vec u} \times {\vec B} - \eta \mu_{0}{\vec J}\;, \\
   \frac{\partial \ln \rho}{\partial t} &=& - ({\vec u} \cdot \nabla)\ln \rho + \nabla \cdot {\vec u}\;, \\
   \frac{\partial {\vec u}}{\partial t} &=& - ({\vec u} \cdot \nabla){\vec u} - \frac{1}{\rho}\nabla p - 2\,\vec{\Omega} \times {\vec u} + {\vec g} + \nonumber \\
   && \hspace{2.5cm} + \frac{1}{\rho}{\vec J} \times {\vec B} + \frac{1}{\rho} \nabla \cdot \tens{\sigma}\;,\label{equ:momentum}\\
   \frac{\partial e}{\partial t} &=& - ({\vec u} \cdot \nabla)e -
\frac{p}{\rho}(\nabla \cdot {\vec u}) + \frac{1}{\rho}\nabla \cdot (\chi \rho
\nabla e) + \nonumber \label{equ:ee} \\ 
   && \hspace{2.5cm} + \Gamma_{\rm visc} + \Gamma_{\rm Joule} - \Gamma_{\rm cool}\;,
   \end{eqnarray}
   where ${\vec A}$ is the vector potential, $\vec{u}$ the velocity,
   ${\vec B} = \nabla \times {\vec A}$ the magnetic field, ${\vec J} =
   \nabla \times {\vec B}/\mu_{0} $ the current density, $\rho$ the
   mass density, $p$ the pressure, ${\vec g} = g{\vec{\hat{e}}}_{z}$
   the (constant) gravity, and $e = c_{\rm V} T$ the internal energy
   per unit mass. $\eta$ and $\nu$ are the magnetic diffusivity and
   kinematic viscosity, respectively, $\mu_{0}$ the vacuum
   permeability, and $\chi$ the thermal diffusivity. The equation of
   state is that of an ideal gas
   \begin{eqnarray}
   p = \rho e (\gamma - 1)\;,
   \end{eqnarray}
    where the ratio of the specific heats $\gamma = c_{\rm P}/c_{\rm
   V} = 5/3$. $\tens{\sigma} = 2\rho \nu \tens{S}$ is the stress
   tensor where
   \begin{eqnarray}
   S_{ij} = \frac{1}{2} \Big(\frac{\partial u_{i}}{\partial x_{j}}
+ \frac{\partial u_{j}}{\partial x_{i}} - \frac{2}{3}\,\delta_{ij} \nabla \cdot {\vec u} \Big)\;.
   \end{eqnarray}
   The terms responsible for viscous and Joule heating can be written
   as
   \begin{eqnarray}
   \Gamma_{\rm visc} &=&  2\nu S_{ij} \frac{\partial u_{i}}{\partial x_{j}} \;, \\ 
   \Gamma_{\rm Joule}&=&  \frac{\eta \mu_{0}}{\rho} {\vec J}^{2}\;.
   \end{eqnarray}
   We use a narrow cooling layer on top of the convection zone, 
   cooled with a term
   \begin{eqnarray}
   \Gamma_{\rm cool} = \frac{1}{t_{\rm cool}} f(z) (e - e_{0})\;,
   \label{equ:cool}
   \end{eqnarray}
   where $t_{\rm cool}$ is a cooling time, chosen to be short enough
   for the upper boundary to stay isothermal, $f(z)$ a function which
   vanishes everywhere else but in the interval $z_{0} \le z < z_{1}$,
   and $e_{0} = e(z_{0})$ the value of internal energy at the top of 
   the box. This parametrisation mimics the radiative losses at the
   stellar surface and, although still rather simple in comparison to
   the real surface layers of stars, works as a more realistic boundary 
   condition and stabilises the numerics better than just imposing a 
   constant temperature at the boundary.

   We adopt periodic boundary conditions in the horizontal directions,
   and closed stress free boundaries at the top and bottom. The
   temperature is kept fixed at the top of the box and a constant heat
   flux is applied at the bottom
   \begin{eqnarray}
   \frac{\partial u_{x}}{\partial z} = \frac{\partial u_{y}}{\partial z} = u_{z} &=& 0\; \hspace{2cm} {\rm at} \;\;\;z = z_{0},z_{3}\;; \\
   e(z_{0}) &=& e_{0}\;, \\
   \frac{\partial e}{\partial z}\Big|_{z_{3}} &=& \frac{g}{(\gamma - 1)(m_{3} + 1)}\;,
   \label{equ:dez}
   \end{eqnarray}
   where $m_{3}$ is the polytropic index of the lower overshoot layer.

   We adopt the same dimensionless quantities and box dimensions as 
   Brandenburg et al. (\cite{Brand96}), and Ossendrijver et al. 
   (\cite{Osse01}, \cite{Osse02}). Thus by setting
   \begin{eqnarray*}
   d = \rho_{0} = g = \mu_{0} = c_{\rm P} = 1\;,
   \end{eqnarray*}
   length is measured with respect to the depth of the unstable layer,
   $d = z_{2} - z_{1}$, and density in units of the initial value at
   the bottom of the convectively unstable layer, $\rho_{0}$. Time is
   measured in units of the free fall time, $\sqrt{d/g}$, velocity in
   units of $\sqrt{dg}$, magnetic field in terms of
   $\sqrt{d\rho_{0}\mu_{0}g}$, and entropy in terms of $c_{\rm P}$.
   The box has horizontal dimensions $L_{x} = L_{y} = 4$, and
   $L_{z} = 2$ in the vertical direction. The vertical coordinates are
   set to be ($z_{0}, z_{1}, z_{2}, z_{3}$) = ($-0.15, 0, 1, 1.85$),
   where $z_{0}$ and $z_{3}$ are the top and bottom of the box, and
   $z_{1}$ and $z_{2}$ are the boundaries between stable and unstable
   layers.

   The dimensionless parameters controlling the calculations are the
   kinematic and magnetic Prandtl numbers Pr and Pm, and the Taylor,
   Rayleigh, and Chandrasekhar numbers, denoted by Ta, Ra, and Ch,
   respectively. The relative importances of thermal and magnetic
   diffusion against the kinetic one are measured by the kinetic and
   magnetic Prandtl numbers
   \begin{eqnarray}
     {\rm Pr} &=& \frac{\nu}{\chi_{0}}\;, \\
     {\rm Pm} &=& \frac{\nu}{\eta}\;,
   \end{eqnarray}
   where $\chi_{0}$ is the reference value of the thermal diffusivity,
   taken from the middle of the unstably stratified layer. In the
   present study, $\nu$ and $\eta$ are constants. We set Pr = 0.4
   which is a compromise between computational time (large diffusivity
   enforces a smaller time step) and realistic physics (in the Sun Pr
   $\ll 1$). In the magnetic runs we set Pm = 1 because of a similar
   compromise between realistic physics and the need to keep Rm
   reasonably large.

   Rotation is measured by the Taylor number
   \begin{eqnarray}
     {\rm Ta} = \Big( \frac{2\Omega d^{2}}{\nu} \Big)^{2}\;.
   \end{eqnarray}
   A related dimensionless quantity is the Coriolis number, which is
   twice the inverse of the Rossby number, Co = $2\,\Omega \tau$, where
   $\tau = d/u_{t}$ is the convective turnover time, and $u_{t} =
   \langle \vec{u}'^{2} \rangle^{1/2}_{V}$ the rms-value of velocity
   fluctuations averaged over the convectively unstable layer and
   time. In our calculations Co varies about two orders of magnitude
   from about 0.1 to roughly 15 (see Tables \ref{tab:Calcu} and 
   \ref{tab:HiresCalcu}).

   Convection efficiency is measured by the Rayleigh number
   \begin{eqnarray}
     {\rm Ra} = \frac{d^{4}g\delta}{\chi_{0}\nu H_{\rm p}}\;,
   \end{eqnarray}
   where $\delta = \nabla - \nabla_{\rm ad}$ is the superadiabaticity,
   measured as the difference between the radiative and the adiabatic
   logarithmic temperature gradients, and $H_{\rm p}$ the pressure
   scale height, both evaluated in the middle of the unstably
   stratified layer in the non-convecting hydrostatic reference
   solution. The Schwarzschild criterion states that Ra $> 0$ for the
   convective instability to occur but in reality it must exceed some
   threshold value $\rm Ra_{c}$. Rotation is known to increase $\rm
   Ra_{c}$ in general (Chandrasekhar \cite{Chandra61}) and its value
   at different latitudes varies (e.g. Hathaway \& Somerville
   \cite{Hatha83}). The Rayleigh numbers ($2.5 \cdot 10^{5}$ and
   $10^{6}$) used in this study are amply supercritical.

   The magnetic field strength is given by the Chandrasekhar number
   \begin{eqnarray}
     {\rm Ch} = \frac{\mu_{0} B_{0}^{2} d^{2}}{4 \pi \rho_{0} \nu \eta}\;,
   \end{eqnarray}
   where $B_{0}$ is the magnitude of the imposed field. We apply 
   pseudo-vacuum boundary conditions
   \begin{eqnarray}
     B_{x} = B_{y} = \frac{\partial B_{z}}{\partial z} = 0\;.
   \end{eqnarray}

   We define the Reynolds numbers as
   \begin{eqnarray}
     {\rm Re} = \frac{u_{t} d}{\nu}\;,\\
     {\rm Rm} = \frac{u_{t} d}{\eta}\;.
   \end{eqnarray}

   The initial stratification is polytropic, described by the indices
   $m_{1},\ m_{2}$, and $m_{3}$ for the three layers. We set $m_{1} =
   \infty,\ m_{2} = 1$, and $m_{3} = 3$, respectively, which means
   that the cooling layer is initially isothermal, and the
   stratification of the lower overshoot layer resembles that of the
   solar model of Stix (\cite{Stix02}). For the
   convective instability to occur, the Rayleigh number must be
   positive, which requires that $m_{1} < m_{\rm ad} = 3/2$. The
   stratification for the internal energy $e$ can thus be written as
   \begin{eqnarray} \label{equ:e} 
      e(z_{0} \leq z \leq z_{1}) & = & e_{0}\;,\nonumber \\ e(z_{1} < z \leq z_{2}) & = & e_{1} + \frac{g\,(z - z_{1})}{(\gamma - 1)(m_{2} + 1)}\;, \\ e(z_{2} < z \leq z_{3}) & = & e_{2} + \frac{g\,(z - z_{2})}{(\gamma - 1)(m_{3}
   + 1)}\;. \nonumber 
   \end{eqnarray} 
   Corresponding equations for the density are 
   \begin{eqnarray} 
      \rho(z_{0} \leq z \leq z_{1}) & = & \rho_{0} \exp \bigg\{ \frac{g\,(z - z_{0})}{(\gamma - 1)e_{0}} \bigg\}\;,\nonumber \\ 
      \rho(z_{1} < z \leq z_{2}) & = & \rho_{1} \bigg[ 1 + \frac{g\,(z - z_{1})}{(\gamma - 1)(m_{2} + 1)\,e_{1}} \bigg]^{m_{2}}\;, \\ 
      \rho(z_{2} < z \leq z_{3}) & = & \rho_{2} \bigg[ 1 + \frac{g\,(z - z_{2})}{(\gamma - 1)(m_{3} + 1)\,e_{2}} \bigg]^{m_{3}}\;. \nonumber 
   \end{eqnarray}

   Initially the radiative flux, $\vec{F}_{\rm rad} = \kappa \nabla
   e$, where $\kappa = \gamma \rho \chi$ is the thermal conductivity,
   carries all of the energy through the domain. This constraint
   together with the Eqs.~(\ref{equ:e}) define the thermal
   conductivities in each layer as
   \begin{eqnarray}
     \frac{\kappa_{i}}{\kappa_{j}} = \frac{m_{j} + 1}{m_{i} + 1}\;,
   \end{eqnarray}
   where $m_{i}$ and $m_{j}$ are the polytropic indices of the
   respective layers. In the calculations the vertical profile of
   $\kappa$ is kept constant, which with the boundary condition for
   the internal energy, Eq.~(\ref{equ:dez}), assures that the heat
   flux through the domain is constant at all times. Furthermore,
   $\kappa$ is smoothed at the interfaces of the layers over an
   interval of about $0.1d$ by a sine function where the inflection
   point occurs at the gridpoint nearest to the interface. The initial
   state is perturbed with small scale velocity fluctuations of the
   order of $10^{-2} \sqrt{dg}$ which are deposited in the
   convectively unstable layer.

   The code used in the calculations is a modified version of that
   presented in Caunt \& Korpi (\cite{CauKo}). The numerical method is
   based on sixth order accurate explicit spatial discretisation and a
   third order accurate Adams-Bashforth-Moulton predictor-corrector
   time stepping scheme. The code is parallelised using message
   passing interface (MPI). The calculations were carried out on the
   IBM eServer Cluster 1600 supercomputer hosted by CSC Scientific
   Computing Ltd., in Espoo, Finland.

   \begin{table*}
   \centering
      \caption[]{Summary of the Ra = $2.5 \cdot 10^{5}$ calculations. In all
      cases investigated Pr = 0.4, $\delta = 0.1$,
      $H_{\rm p} = 0.45$, and $e_{0} = 0.3$. With these parameters
      $\nu = 5.96 \cdot 10^{-4}$. The full computational domain
      extends over about $4.2$ pressure and $2.7$ density scale
      heights. The averages are calculated over the unstable layer and
      time.}
      \vspace{-0.75cm}
      \label{tab:Calcu}
     $$
         \begin{array}{p{0.08\linewidth}cccccccccrc}
            \hline
            \noalign{\smallskip}
            Run      & \Theta & {\rm Re} & {\rm Ta} & {\rm Co} & \langle
\vec{u}^{2} \rangle^{1/2}_{V} & \langle \vec{u}'^{2} \rangle^{1/2}_{V} &
\langle u_{x}'^{2} \rangle^{1/2}_{V} & \langle u_{y}'^{2} \rangle^{1/2}_{V} & \langle u_{z}'^{2} \rangle^{1/2}_{V} & \mathcal{H}[10^{-4}] & \Delta T/\tau\\
            \noalign{\smallskip}
            \hline
            \noalign{\smallskip}
            Co01-00   &  \;\;\; 0 \degr & 143 & 203         & 0.10 & 0.085 & 0.085 & 0.041 & 0.041 & 0.061 & -3.8 & 103 \\
	    Co01-7.5  & -7.5\degr & 143 & 203               & 0.10 & 0.086 & 0.085 & 0.041 & 0.041 & 0.062 & -6.8 & 104 \\
	    Co01-15   & -15 \degr & 143 & 203               & 0.10 & 0.085 & 0.085 & 0.041 & 0.041 & 0.061 & -1.3 & 103 \\
	    Co01-30   & -30 \degr & 142 & 203               & 0.10 & 0.085 & 0.085 & 0.041 & 0.041 & 0.061 &  1.6 & 103 \\
	    Co01-45   & -45 \degr & 143 & 203               & 0.10 & 0.086 & 0.085 & 0.041 & 0.042 & 0.061 &  3.1 & 102 \\
	    Co01-60   & -60 \degr & 143 & 203               & 0.10 & 0.085 & 0.085 & 0.042 & 0.041 & 0.062 &  7.3 & 103 \\
	    Co01-75   & -75 \degr & 143 & 203               & 0.10 & 0.085 & 0.085 & 0.041 & 0.042 & 0.061 &  5.1 & 103 \\
	    Co01-82.5 &-82.5\degr & 144 & 203               & 0.10 & 0.086 & 0.086 & 0.042 & 0.042 & 0.061 & -2.2 & 104 \\
            \hline
            \noalign{\smallskip}
	    Co05-7.5  & -7.5\degr & 144 & 5.08 \cdot 10^{3} & 0.50 & 0.088 & 0.086 & 0.042 & 0.042 & 0.061 &  0.0 & 106 \\
	    Co05-30   & -30 \degr & 144 & 5.08 \cdot 10^{3} & 0.50 & 0.087 & 0.086 & 0.042 & 0.042 & 0.061 &  8.7 & 106 \\
	    Co05-60   & -60 \degr & 145 & 5.08 \cdot 10^{3} & 0.50 & 0.087 & 0.086 & 0.043 & 0.043 & 0.061 &  3.6 & 107 \\
	    Co05-82.5 &-82.5\degr & 144 & 5.08 \cdot 10^{3} & 0.50 & 0.086 & 0.086 & 0.042 & 0.043 & 0.061 &  8.6 & 106 \\
            \hline
            \noalign{\smallskip}
	    Co1-00    & \;\;\;  0 \degr & 143 & 2.03 \cdot 10^{4} & 1.00 & 0.093 & 0.085 & 0.042 & 0.041 & 0.061 & -5.8 & 105 \\
	    Co1-7.5   & -7.5\degr & 146 & 2.03 \cdot 10^{4} & 0.99 & 0.092 & 0.087 & 0.043 & 0.043 & 0.061 & 13.8 & 107 \\
	    Co1-15    & -15 \degr & 145 & 2.03 \cdot 10^{4} & 0.99 & 0.089 & 0.087 & 0.041 & 0.044 & 0.061 & 15.5 & 107 \\
	    Co1-30    & -30 \degr & 146 & 2.03 \cdot 10^{4} & 0.99 & 0.089 & 0.087 & 0.043 & 0.045 & 0.060 & 18.2 & 108 \\
	    Co1-45    & -45 \degr & 146 & 2.03 \cdot 10^{4} & 0.99 & 0.088 & 0.087 & 0.043 & 0.045 & 0.060 & 22.9 & 107 \\
	    Co1-60    & -60 \degr & 144 & 2.03 \cdot 10^{4} & 1.01 & 0.086 & 0.086 & 0.043 & 0.044 & 0.059 & 25.5 & 105 \\
	    Co1-75    & -75 \degr & 142 & 2.03 \cdot 10^{4} & 1.03 & 0.085 & 0.085 & 0.043 & 0.043 & 0.059 & 23.4 & 104 \\
	    Co1-82.5  &-82.5\degr & 142 & 2.03 \cdot 10^{4} & 1.03 & 0.085 & 0.085 & 0.042 & 0.042 & 0.059 & 26.6 & 104 \\
            \hline
            \noalign{\smallskip}
	    Co2-7.5   & -7.5\degr & 150 & 8.14 \cdot 10^{4} & 1.93 & 0.095 & 0.089 & 0.042 & 0.050 & 0.060 &  2.4 & 110 \\
	    Co2-30    & -30 \degr & 148 & 8.14 \cdot 10^{4} & 1.97 & 0.090 & 0.088 & 0.046 & 0.050 & 0.057 & 32.4 & 109 \\
	    Co2-60    & -60 \degr & 142 & 8.14 \cdot 10^{4} & 2.06 & 0.085 & 0.085 & 0.044 & 0.046 & 0.056 & 52.8 & 105 \\
	    Co2-82.5  &-82.5\degr & 136 & 8.14 \cdot 10^{4} & 2.17 & 0.081 & 0.081 & 0.041 & 0.041 & 0.056 & 41.5 & 100 \\
            \hline
            \noalign{\smallskip}
	    Co4-00h   & \;\;\;  0 \degr & 181 & 3.25 \cdot 10^{5} & 3.23 & 0.111 & 0.108 & 0.038 & 0.073 & 0.066 &  9.7 & 137 \\
	    Co4-00l   & \;\;\;  0 \degr & 170 & 3.25 \cdot 10^{5} & 3.40 & 0.103 & 0.101 & 0.035 & 0.067 & 0.064 & 16.8 & 17.0 \\
	    Co4-7.5h  & -7.5\degr & 167 & 3.25 \cdot 10^{5} & 3.48 & 0.101 & 0.100 & 0.047 & 0.062 & 0.061 & 48.7 & 127 \\
	    Co4-7.5l  & -7.5\degr & 167 & 3.25 \cdot 10^{5} & 3.50 & 0.100 & 0.099 & 0.048 & 0.063 & 0.060 & 31.7 & 52.0 \\
	    Co4-15l   & -15 \degr & 157 & 3.25 \cdot 10^{5} & 3.71 & 0.094 & 0.094 & 0.050 & 0.056 & 0.056 &-11.7 & 120 \\
	    Co4-30h   & -30 \degr & 146 & 3.25 \cdot 10^{5} & 4.03 & 0.087 & 0.087 & 0.048 & 0.050 & 0.052 & 11.4 & 111 \\
	    Co4-30l   & -30 \degr & 146 & 3.25 \cdot 10^{5} & 4.02 & 0.087 & 0.087 & 0.048 & 0.051 & 0.051 &  7.2 & 111 \\
	    Co4-45l   & -45 \degr & 135 & 3.25 \cdot 10^{5} & 4.39 & 0.081 & 0.081 & 0.045 & 0.045 & 0.049 & 28.7 & 103 \\
	    Co4-60h   & -60 \degr & 129 & 3.25 \cdot 10^{5} & 4.62 & 0.077 & 0.077 & 0.042 & 0.041 & 0.050 & 30.8 & 96 \\
	    Co4-60l   & -60 \degr & 132 & 3.25 \cdot 10^{5} & 4.54 & 0.079 & 0.079 & 0.043 & 0.042 & 0.050 & 26.7 & 100 \\
	    Co4-75l   & -75 \degr & 127 & 3.25 \cdot 10^{5} & 4.71 & 0.076 & 0.076 & 0.039 & 0.040 & 0.051 & 28.8 & 96.3 \\
	    Co4-82.5h &-82.5\degr & 126 & 3.25 \cdot 10^{5} & 4.75 & 0.075 & 0.075 & 0.039 & 0.039 & 0.051 & 32.7 & 95.7 \\
	    Co4-82.5l &-82.5\degr & 128 & 3.25 \cdot 10^{5} & 4.69 & 0.076 & 0.076 & 0.040 & 0.040 & 0.051 & 30.4 & 93.9 \\
            \hline
            \noalign{\smallskip}
	    Co7-7.5   & -7.5\degr & 173 & 9.97 \cdot 10^{5} & 5.98 & 0.105 & 0.103 & 0.053 & 0.060 & 0.063 &  134 & 131 \\
	    Co7-30    & -30 \degr & 142 & 9.97 \cdot 10^{5} & 7.26 & 0.085 & 0.084 & 0.050 & 0.048 & 0.048 & -25.4& 107 \\
	    Co7-60    & -60 \degr & 123 & 9.97 \cdot 10^{5} & 8.51 & 0.074 & 0.074 & 0.041 & 0.039 & 0.046 & -4.5 & 89.9 \\
	    Co7-82.5  &-82.5\degr & 123 & 9.97 \cdot 10^{5} & 8.60 & 0.074 & 0.073 & 0.039 & 0.039 & 0.048 &  1.1 & 89.1 \\
            \hline
            \noalign{\smallskip}
	    Co10-00   & \;\;\;  0 \degr & 321 & 2.03 \cdot 10^{6} & 4.81 & 0.248 & 0.191 & 0.034 & 0.142 & 0.112 & -18.3 & 238 \\
            Co10-7.5  & -7.5\degr & 172 & 2.03 \cdot 10^{6} & 8.65 & 0.104 & 0.102 & 0.055 & 0.059 & 0.062 & 93.9 & 131 \\
	    Co10-15   & -15 \degr & 148 & 2.03 \cdot 10^{6} & 9.94 & 0.089 & 0.088 & 0.047 & 0.049 & 0.055 & -57.1& 33.8 \\
	    Co10-30   & -30 \degr & 136 & 2.03 \cdot 10^{6} & 10.9 & 0.081 & 0.081 & 0.047 & 0.044 & 0.049 & -46.4& 90.2 \\
	    Co10-45   & -45 \degr & 117 & 2.03 \cdot 10^{6} & 12.7 & 0.070 & 0.070 & 0.041 & 0.037 & 0.042 & -33.3& 75.6 \\
	    Co10-60   & -60 \degr & 112 & 2.03 \cdot 10^{6} & 13.3 & 0.067 & 0.067 & 0.038 & 0.036 & 0.041 & -20.0& 76.2 \\
	    Co10-75   & -75 \degr &  96 & 2.03 \cdot 10^{6} & 15.7 & 0.058 & 0.058 & 0.031 & 0.029 & 0.038 & -13.3& 22.6  \\
	    Co10-82.5 &-82.5\degr & 118 & 2.03 \cdot 10^{6} & 12.5 & 0.071 & 0.070 & 0.041 & 0.038 & 0.041 & -25.2& 79.0 \\
            \hline
         \end{array}
     $$ 
   \vspace{-0.6cm}
   \end{table*}

   \begin{table*}
   \centering
      \caption[]{Summary of the Ra = $10^{6}$ calculations. In these
      runs $\nu = 2.98 \cdot 10^{-4}$, other parameters as indicated
      in Table \ref{tab:Calcu}.}
      \vspace{-0.75cm}
      \label{tab:HiresCalcu}
     $$
         \begin{array}{p{0.1\linewidth}cccccccccrc}
            \hline
            \noalign{\smallskip}
            Run      & \Theta & {\rm Re} & {\rm Ta} & {\rm Co} & \langle
\vec{u}^{2} \rangle^{1/2}_{V} & \langle \vec{u}'^{2} \rangle^{1/2}_{V} &
\langle u_{x}'^{2} \rangle^{1/2}_{V} & \langle u_{y}'^{2}
\rangle^{1/2}_{V} & \langle u_{z}'^{2} \rangle^{1/2}_{V} & \mathcal{H}[10^{-4}] & \Delta T/\tau\\
            \noalign{\smallskip}
            \hline
            \noalign{\smallskip}
	    HICo1-7.5   &  -7.5 \degr      & 263 & 5.79 \cdot 10^{4} & 0.92 & 0.081 & 0.078 & 0.038 & 0.039 & 0.056 &  5.1 & 45.9\\
	    HICo1-30    &   -30 \degr      & 266 & 5.79 \cdot 10^{4} & 0.92 & 0.081 & 0.079 & 0.040 & 0.041 & 0.055 &  9.8 & 46.5\\
	    HICo1-60    &   -60 \degr      & 269 & 5.79 \cdot 10^{4} & 0.91 & 0.080 & 0.080 & 0.041 & 0.041 & 0.055 & 15.6 & 46.9\\
	    HICo1-82.5  & -82.5 \degr      & 265 & 5.79 \cdot 10^{4} & 0.93 & 0.079 & 0.079 & 0.040 & 0.040 & 0.054 & 11.0 & 46.3\\
            \hline
            \noalign{\smallskip}
	    HICo4-7.5   &  -7.5 \degr      & 301 & 1.12 \cdot 10^{6} & 3.60 & 0.090 & 0.090 & 0.045 & 0.054 & 0.055 & -7.8 & 52.5\\
	    HICo4-30    &   -30 \degr      & 267 & 1.12 \cdot 10^{6} & 4.11 & 0.080 & 0.080 & 0.044 & 0.046 & 0.047 &  6.9 & 46.6\\
	    HICo4-60    &   -60 \degr      & 245 & 1.12 \cdot 10^{6} & 4.55 & 0.073 & 0.073 & 0.039 & 0.039 & 0.047 & 26.7 & 42.7\\
	    HICo4-82.5  & -82.5 \degr      & 237 & 1.12 \cdot 10^{6} & 4.71 & 0.071 & 0.071 & 0.037 & 0.037 & 0.048 & 28.8 & 41.3\\
            \hline
            \noalign{\smallskip}
	    HICo10-7.5  &  -7.5 \degr      & 328 & 7.03 \cdot 10^{6} & 8.38 & 0.098 & 0.098 & 0.053 & 0.056 & 0.059 & 94.0 & 41.7\\
	    HICo10-30   &   -30 \degr      & 241 & 7.03 \cdot 10^{6} & 11.4 & 0.072 & 0.072 & 0.042 & 0.040 & 0.041 & -38.5& 35.1\\
	    HICo10-60   &   -60 \degr      & 205 & 7.03 \cdot 10^{6} & 13.7 & 0.061 & 0.061 & 0.034 & 0.032 & 0.039 & -13.6& 30.0\\
	    HICo10-82.5 & -82.5 \degr      & 201 & 7.03 \cdot 10^{6} & 14.0 & 0.060 & 0.060 & 0.032 & 0.031 & 0.040 & -8.4 & 28.4\\
            \hline
         \end{array}
     $$ 
   \end{table*}


\section{Summary of the computations, tests, and methods}
\label{sec:cameth}
   The parameters used in the calculations are listed in Tables
   \ref{tab:Calcu} and \ref{tab:HiresCalcu} along with a few
   diagnostic quantities calculated from the data. The calculations
   produced are labelled with the Coriolis number, estimated using the
   turbulent velocity from a non-rotating calculation, and latitude;
   as the turbulent velocity is changing as function of latitude and
   rotation especially in the rapid rotation regime, the actual
   Coriolis number realised in a calculation does not exactly coincide
   with its label (see the fifth column). The investigated Coriolis
   numbers vary between 0.1 and 15; calculations with even higher
   values would have required higher spatial resolution (see below).

   To cut the computational expense some compromise concerning the
   latitudinal coverage had to be made. In sets Co01, Co1, Co4, and
   Co10 calculations were made at latitudes $0, -7.5\degr, -15\degr,
   -30\degr, - 45\degr, -60\degr, -75\degr$, and $-82.5\degr$, whereas
   in the sets Co05, Co2, and Co7, only at latitudes $-7.5\degr,
   -30\degr, -60\degr$, and $-82.5\degr$. On the poles, the Reynolds
   stresses should vanish due to symmetry and thus the runs there were
   limited to a few test cases to confirm the expected behaviour.

   The numerical resolution was kept fixed at a rather moderate $64^{3}$
   grid points for the sets Co01, Co05, Co1, and Co2. This resolution
   turned out to be insufficient for the rapid rotation cases where
   strong azimuthal mean flows were generated. Since the spatial
   resolution in the vertical direction was already higher than in the
   horizontal directions and the vertical velocity was less affected by
   rotation (compare columns 9 and 10 in Tables \ref{tab:Calcu} and
   \ref{tab:HiresCalcu}), increasing the horizontal resolution
   moderately was a sufficient solution. Thus, for the full sets Co7,
   and Co10, and a subset of the Co4 runs a resolution of $96^{2} \times
   64$ was used. Since one calculation in the Co10 set still ended in a
   numerical instability (Co10-15), we decided not to increase rotation
   further with the expense of even higher computational needs in the
   present study. Comparison with the two resolutions can be made within
   the Co4 set. The different resolutions are labelled with the suffix l
   (for lower resolution) and h (for higher resolution) in Table
   \ref{tab:Calcu}. The correspondance of the calculated quantities is
   satisfactory between the low and high resolution cases.  We feel that
   the results obtained this way are consistent.

   To see how the turbulent dynamics depends on Ra, we have performed
   subsets (latitudes $-7.5\degr, -30\degr, -60\degr$, and
   $-82.5\degr$) of sets Co1, Co4, and Co10 with a Rayleigh number of
   $10^{6}$. The numerical resolution was $128^{3}$ for the HICo1 and
   HICo4 sets and $192^{2} \times 128$ for the set HICo10. See Table
   \ref{tab:HiresCalcu} for further details. In what follows, we will
   present results mostly from the Ra = $2.5 \cdot 10^{5}$ runs and
   report on the differences between the two Rayleigh numbers when
   they are significant.

\subsection{Mean and fluctuating quantities}

   When measuring the turbulent properties of a flow one must make a
   clear distinction between the mean and fluctuating parts of the
   quantities involved. In this paper we use the following definitions:
   the mean of a quantity $q_{i}$ is taken to be the horizontal average
   \begin{eqnarray}
     \langle q_{i} \rangle = \frac{1}{L_{x}L_{y}}\int\!\!\int q_{i} dx dy\;,
   \end{eqnarray}
   and the fluctuation is
   \begin{eqnarray}
     q'_{i} = q_{i} - \langle q_{i} \rangle\;,
   \end{eqnarray}
   where $q_{i}$ is the total value of the quantity. Using the
   Reynolds rules, the mean value of a correlation of two fluctuating
   quantities $q'_{i}q'_{j}$ can be written as
   \begin{eqnarray}
     \langle q'_{i}q'_{j} \rangle = \langle q_{i}q_{j} \rangle - \langle q_{i} \rangle \langle q_{j} \rangle\;,
     \label{equ:corr}
   \end{eqnarray}
   from which one can deduce the rms value of the fluctuating quantity
   as
   \begin{eqnarray}
     \langle q_{i}'^{2} \rangle^{1/2} = \sqrt{ \langle q_{i}^{2} \rangle - \langle q_{i} \rangle^{2} }\;.
   \end{eqnarray}
   Furthermore, the volume average of quantity $q_{i}$ over the unstable
   layer is defined by
   \begin{eqnarray}
     \langle q_{i} \rangle_{V} = \frac{1}{z_{2}-z_{1}} \int_{z_{1}}^{z_{2}} \langle q_{i} \rangle dz\;.
   \end{eqnarray}
   Additionally, a time average is defined by
   \begin{eqnarray}
     \langle q_{i} \rangle_{t} = \frac{1}{\Delta T} \int_{t_{0}}^{t_{1}} \langle q_{i} \rangle dt\;,
   \end{eqnarray}
   where $\Delta T$ is the length of the average. The Reynolds rules
   are only approximately valid for time averages. However, the longer
   the averaging interval, the better the approximation. In the
   present study the time averaging intervals are long enough for the
   Reynolds rules to be very nearly obeyed. All the calculations were
   started from the same initial convectively unstable state from
   which it in most cases takes about 50 time units for the convection
   to reach a saturation. In order to remove the effects of the
   initial transient we always neglect the first 100 time units from
   the calculation of the time average. For rapid rotation sets Co7
   and Co10 with the lower Rayleigh number, the transient is somewhat
   longer which is taken into account in the calculation of the
   averages. In what follows, the time average is applied always in
   addition to a horizontal or volume average.

\subsection{Statistics and error estimation}

\begin{figure}
   \centering
   \includegraphics[width=0.5\textwidth]{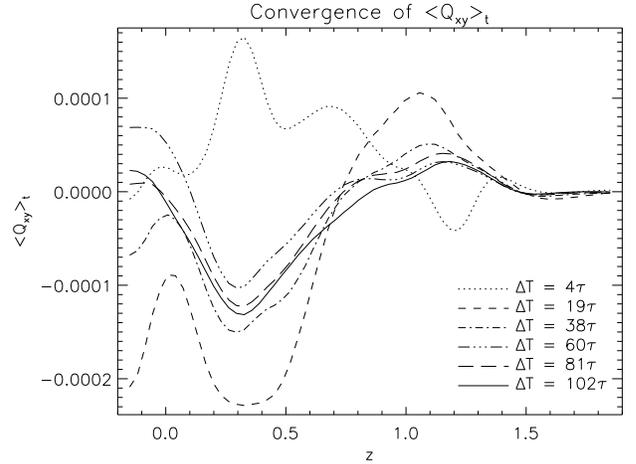}
      \caption{Convergence of the Reynolds stress component $\langle
      Q_{xy} \rangle$ from the run Co1-45. The approximate number of
      turnover times in the average of the corresponding linestyle is
      indicated in the plot.}
       \label{pic:convergence}
   \end{figure}

   It is known from previous studies (e.g. Pulkkinen et
   al. \cite{Pulkki93}; Chan \cite{Chan01}) that the Reynolds stresses
   are highly fluctuating quantities and thus large data sets and as
   much averaging as possible is needed to derive reliable
   results. Chan (\cite{Chan01}) estimates that reasonable convergence
   of the Reynolds stresses is obtained by averaging at least over a
   hundred turnover times. Fig.~\ref{pic:convergence} shows the
   horizontally-averaged Reynolds stress component $\langle Q_{xy}
   \rangle$ from the run Co1-45, averaged over six different time
   intervals, corresponding to 4, 19, 38, 60, 81, and 102 convective
   turnover times. The convergence of this quantity is indeed quite
   slow in this case, but the main features of the stress do not
   change significantly if the average is taken over more than 40
   turnover times.

   We list the integration times (minus the initial transient) of the
   calculations in Tables \ref{tab:Calcu} and \ref{tab:HiresCalcu},
   from which we see that the integration time for the bulk of the
   runs exceed these limits, except for the cases which were
   terminated due to numerical problems. In order to restrict the
   computational costs, the $\rm Ra = 10^{6}$ runs were integrated
   only approximately half the time (in dimensionless units) in
   comparison to the low resolution calculations.

   We estimate the error of the averaged quantities using a modified
   mean error of the mean, defined as
   \begin{eqnarray}
     \varepsilon = \bigg[ \frac{\tau}{\Delta T} \frac{1}{(N - 1)} \sum_{i = 1}^N (f(i) - \langle f \rangle)^{2} \bigg]^{1/2}\;, 
     \label{equ:error}
   \end{eqnarray}
   where $\frac{\Delta T}{\tau}$ is the number of turnover times in
   the average, giving an estimate of the number of independent
   realisations of convection during a calculation, and $N$ the
   number of data points in the average $\langle f \rangle$.

\section{The $\Lambda$-effect and turbulent heat transport}
\label{sec:theLaTH}

   \subsection{$\Lambda$-effect}

   We begin by introducing the mean
   field momentum equation (the Reynolds equation) in the
   magnetohydrodynamical regime. Firstly, the velocity and magnetic
   fields are separated into the mean and fluctuating parts, ${\vec U}
   = \langle {\vec u} \rangle + {\vec u}'$ and ${\vec B} = \langle
   {\vec B} \rangle + {\vec b}'$, which are then inserted into the
   momentum equation, Eq.~(\ref{equ:momentum}). Taking the mean and
   applying the Reynolds rules one arrives at
   \begin{eqnarray}
     \label{equ:Reynolds}
     \rho \Big( \frac{\partial \langle {\vec u} \rangle}{\partial t} +
\langle {\vec u} \rangle \cdot \nabla \langle {\vec u}\rangle \Big) & =
& - \nabla \cdot (\rho \tens{Q} - \tens{M}) + \rho {\vec g} - \nabla p
\\ \nonumber & & + {\langle \vec J} \rangle \times \langle {\vec B} \rangle - \nabla \cdot \langle \sigma \rangle,
   \end{eqnarray}
   where the fluctuations of $\rho$ and $p$ have been ignored. The
   influence of the small-scale turbulence on the large-scale velocity
   comes through the Reynolds stress $Q_{ij} = \langle u'_{i}u'_{j}
   \rangle$ and the Maxwell stress $M_{ij} = \mu_0^{-1} \langle
   b'_{i}b'_{j} \rangle$ tensors, whilst the influence of the
   large-scale magnetic field through the Lorentz force $\langle {\vec
   J} \rangle \times \langle {\vec B} \rangle$, where $\langle {\vec J}
   \rangle =\mu_0^{-1} \nabla \times \langle {\vec B} \rangle$ is the
   mean electric current.

   Furthermore, by adopting a spherical coordinate system,
   assuming that the quantities are axisymmetric due to which defining
   the mean values as azimuthal averages is meaningful, and neglecting
   the molecular viscosity which is known to be small in stellar
   convection zones, one can derive the equation for the
   angular momentum balance
   \begin{eqnarray} 
   \label{equ:angu}
   \frac{\partial}{\partial t} (\rho s^{2} \Omega) + \nabla \cdot (\rho s^{2} 
   \Omega {\vec u}_{m} & + & s \langle \rho u'_{\phi}{\vec u}' \rangle \nonumber \\
 & - & \mu_0^{-1} \langle b'_{\phi}{\vec b}' \rangle ) =  0\;,
   \end{eqnarray} 
   where $s = r \sin \theta$ is the lever arm, ${\vec u}_{m} =
   (\langle u_{r} \rangle, \langle u_{\theta} \rangle,0)$ the
   meridional circulation, and $\langle u_{\phi} \rangle = r \Omega
   \sin \theta$ describes the basic rotation. From this equation it is
   evident that both the meridional circulation and the turbulent
   stresses contribute to the angular momentum transport.

   If the Boussinesq ansatz was adopted for the Reynolds stresses,
   i.e. writing them in the form of a diffusion term $Q_{ij}=-\nu_t
   \big(\frac{\partial \langle u_{i} \rangle}{\partial
   x_{j}}+\frac{\partial \langle u_{j} \rangle}{\partial x_{i}}
   \big)$, they would act to reduce the gradients in the angular
   velocity rather than as a source of differential rotation. This,
   however, leads to a contradiction between the observed Reynolds
   stress component $Q_{\theta \phi}$ from sunspot proper motions
   (Ward \cite{Ward65}; more recently determined by e.g. Pulkkinen \&
   Tuominen \cite{Pulkki98}) and the one calculated from the observed
   latitudinal differential rotation. As proposed by R\"udiger
   (\cite{R77}) in the form of the theory of the $\Lambda$-effect, the
   rotational influence on a turbulent anisotropic flow may give rise
   to additional stresses proportional to $\Omega$ itself, capable of
   driving differential rotation rather than diffusing it. Taking this
   effect into account, also the sunspot measurements can be explained
   (e.g. Tuominen \& R\"udiger \cite{TR89}; Pulkkinen et
   al. \cite{Pulkki93}).

   The theory of the $\Lambda$-effect is derived using the
   mean-field approach for hydrodynamics (see e.g. R89 for
   a thorough discussion). In a statistically steady and
   non-magnetic medium, the Reynolds stresses may be expanded in
   increasing powers of spatial derivatives of the mean velocities
   \begin{eqnarray} 
     Q_{ij} = \Lambda_{ijk} \Omega_{k}
        -\mathcal{N}_{ijkl} \frac{\partial \langle u_{k} \rangle}{\partial
        x_{l}} + \ldots \;, \label{equ:qlamdif} 
   \end{eqnarray} 
   where $\Lambda_{ijk}$ and $\mathcal{N}_{ijkl}$ are tensors which
   contain the non-diffusive and diffusive parts of the stresses,
   respectively. Higher order derivatives of the mean velocity can be
   neglected under the assumption of scale separation (i.e. that the
   mean-field approach is applicable). In the present study we confine
   our interest to the non-diffusive part of the Reynolds stresses, but
   note here that the diffusive part has been discussed in length by R89
   and more recently investigated with analytical turbulence models
   based on the quasi-linear approach (e.g. Kitchatinov et
   al. \cite{Kitcha94b}).

   An appropriate form of the $\Lambda$-tensor leading to
   non-diffusive expressions can be constructed if the anisotropy due
   to the stratification in the radial direction is taken into account
   (see e.g. R89), resulting in
   \begin{eqnarray}
     Q_{ij} & = &  \left( \begin{array}{ccc}
0 & 0 & \Lambda_V \sin \theta \\
0 & 0 & \Lambda_H \cos \theta \\
\Lambda_V \sin \theta & \Lambda_H \cos \theta & 0 \\
\end{array} \right)
\Omega\;,{\rm where} \label{equ:nondiffrey} \\
     \Lambda_V & = & \nu_t (V^{(0)}(r,\Omega) + V (r,\theta,\Omega)) \label{equ:LambdaV}\;,{\rm and} \\
     \Lambda_H & = & \nu_t H (r,\theta,\Omega) \label{equ:LambdaH}\;.
   \end{eqnarray}
   Here $\nu_t$ is the turbulent viscosity, $\theta$ the colatitude,
   and the coefficients $V$ and $H$ describe the generation of
   vertical (radial) and horizontal (latitudinal) differential
   rotation. In the limit of slow rotation only the fundamental mode
   $V^{(0)}$ is expected to be important, whilst for higher rotation
   rates also the $H$ and higher $V$ modes nonlinear in $\Omega$ may
   become excited (e.g. KR93).

   The $r \theta$-component of the Reynolds stress tensor does not
   enter to the equation of angular momentum transport
   (Eq.~\ref{equ:angu}), but appears in the meridional components of
   the Reynolds equation (Eq.~\ref{equ:Reynolds}), thereby being
   capable of influencing the meridional circulation pattern. As noted
   e.g. by R89, such a correlation may arise from the tensor $\Omega_i
   \Omega_j$, and can therefore be expected to take the form $Q_{r
   \theta} \propto \sin{\theta} \cos{\theta}$. In analogy with the
   Eqs.~(\ref{equ:LambdaV}) and (\ref{equ:LambdaH}) we write
   \begin{eqnarray} 
   Q_{r \theta} & = & \nu_{t} M(r,\theta,\Omega) \sin \theta \cos \theta \;. \label{equ:LambdaM} 
   \end{eqnarray}
   These coefficients are now obtainable from the calculations via the
   equations
   \begin{eqnarray}
     H & = & - \frac{\langle Q_{xy} \rangle_V}{\nu_t \cos{\theta} \Omega} \;, \label{equ:nLambdaH} \\
     V & = & - \frac{\langle Q_{yz} \rangle_V}{\nu_t \sin{\theta} \Omega} \;, \label{equ:nLambdaV} \\
     M & = & \frac{\langle Q_{xz} \rangle_V}{\nu_t \sin{\theta} \cos{\theta} \Omega} \;, \label{equ:nLambdaM}
   \end{eqnarray}
   where the minus signs are due to the transformation from the
   coordinate system used in the calculations to the spherical
   ones.

   The mean velocity field can be afftected by the presence of a
   dynamically significant magnetic field by two principal ways,
   firstly through the large-scale Lorentz force $\mu_0^{-1} \langle
   {\vec J} \rangle \times \langle {\vec B} \rangle$, also known as
   the Malkus-Proctor effect, secondly through the turbulent Maxwell
   stresses $M_{ij}$, both of which are, in general, thought to reduce
   the generation of differential rotation through the
   $\Lambda$-effect; thereby this effect is referred to as the
   $\Lambda$-quenching. In some investigations (e.g. Jennings \& Weiss
   \cite{JW91}) this quenching was assumed to take a similar form to
   the $\alpha$-quenching of mean-field dynamos, namely
   $\frac{\partial \Omega}{\partial r} \propto \frac{\partial
   \Omega_0}{\partial r}/(1+ k B^2)$. One must, however, note that it
   is usually not the differential rotation itself but the
   $\Lambda$-coefficients that are quenched in this simple algebraic
   fashion. More detailed investigations have been elaborated in the
   framework of the quasi-linear approach (e.g. Kitchatinov et
   al. 1994b), indicating that the behaviour of the turbulent stresses
   in the magnetic regime is complex and cannot be described by such a
   simple quenching formula. In contrast, in the regime of a weak
   magnetic field, the stresses can actually be enhanced, although in
   the magnetically dominated regime strong suppression is predicted
   (for more details see Sect.~\ref{subsubsec:maglam}). This effect
   has been taken into account in some solar mean-field dynamo models,
   and it has been found to play an important role, e.g., for
   reproducing the torsional oscillations (K\"uker et
   al. \cite{Kuker96}), minima of magnetic activity reminiscent of the
   observed solar minima (Kichatinov et al. \cite{Kitcha94b}, K\"uker
   et al. \cite{Kuker99}) and longer term variation similar to the
   Gleissberg cycle (Pipin \cite{Pipin99}).

   \subsection{Turbulent heat transport}

   The mean energy equation in terms of the temperature, $T$, can be
   obtained from Eq.~(\ref{equ:ee}) by applying the Reynolds rules
   similarly as above in the case of the momentum equation
   \begin{eqnarray}
     c_{V} \rho \Big( \frac{\partial \langle T \rangle}{\partial t} +
\langle \vec{u} \rangle \cdot \nabla \langle T \rangle \Big) & = & -
\nabla \cdot (\rho c_{V} \langle \vec{u}'T' \rangle) - p \nabla \cdot \langle \vec{u} \rangle \nonumber \\ 
     & & + \nabla (\kappa \nabla \langle T \rangle) + \langle \Gamma \rangle\;,
   \end{eqnarray}
   where $\langle \Gamma \rangle$ contains the contributions from
   viscous and Joule heating and the cooling near the surface. The
   turbulent heat transport is described by the correlation tensor
   $\langle u_{i}'T' \rangle$, which can be expected to 
   be latitude dependent due to the influence of rotation on the
   turbulence. This latitude dependence can induce temperature
   differences which are capable of driving meridional flows, which in
   turn may affect the angular momentum transport (as evident from
   Eq.~\ref{equ:angu}).

   Following the treatise of R89, this tensor can be represented by 
   \begin{eqnarray}
     \langle u_{i}'T' \rangle = \chi_{ij} \delta_{j}\;, \label{equ:deftuhe}
   \end{eqnarray}
   where $\chi_{ij}$ is the eddy heat conductivity tensor and
   $\delta_{j} = - (\nabla_{j}T - g_{j}/c_{\rm P})$ the superadiabatic
   temperature gradient in the direction $j$. In the present model
   geometry the horizontal temperature gradients vanish due to the
   periodic boundaries, and we must thus limit the discussion to the
   meridional components of the tensor. In analogy to the
   $\Lambda$-coefficients, Eqs.~(\ref{equ:LambdaV}) and
   (\ref{equ:LambdaH}), we write $\chi_{ij}$ as
   \begin{eqnarray}
     \chi_{rr} & = & -\chi_{t} VV(r,\theta,\Omega) \;, \label{equ:fitVV} \\
     \chi_{\theta r} & = & -\chi_{t} HV(r,\theta,\Omega) \;, \label{equ:fitHV}
   \end{eqnarray}
   where $\chi_{t}$ is the turbulent heat conductivity for which we
   assume that $\chi_{t} = \nu_{t}$.

   The correlations between velocity and temperature fluctuations
   derived from the calculations can now be used to determine the
   coefficients $VV$ and $HV$ with the equations
   \begin{eqnarray}
     VV & = & \frac{\langle u_{z}'T' \rangle_{V}}{\beta_{z} \chi_{t}}\;, \label{equ:fVV} \\
     HV & = & \frac{\langle u_{x}'T' \rangle_{V}}{\beta_{z} \chi_{t}} \;, \label{equ:fHV}
   \end{eqnarray}
   where the sign changes for both quantities due to the
   transformation to spherical coordinates.


   \section{Results}
   \label{sec:results}

   \begin{figure}
   \centering
      \includegraphics[width=0.5\textwidth]{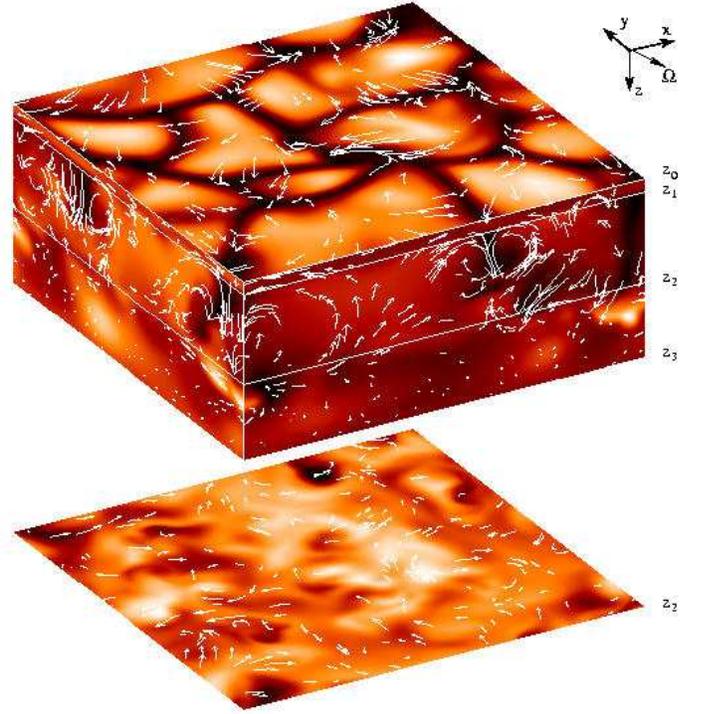}
      \caption{Snapshot of temperature fluctuations and velocity
      vectors from the run HICo1-30. At the top surface of the box we
      show horizontal contours at the $z = z_1$ plane and at the lower
      surface at the $z = z_2$ plane. The coordinate system and the
      direction of the rotation vector are shown in the upper right
      corner.}
      \label{pic:boxplot}
   \end{figure}

   \begin{figure*}
   \centering
      \includegraphics[width=0.95\textwidth]{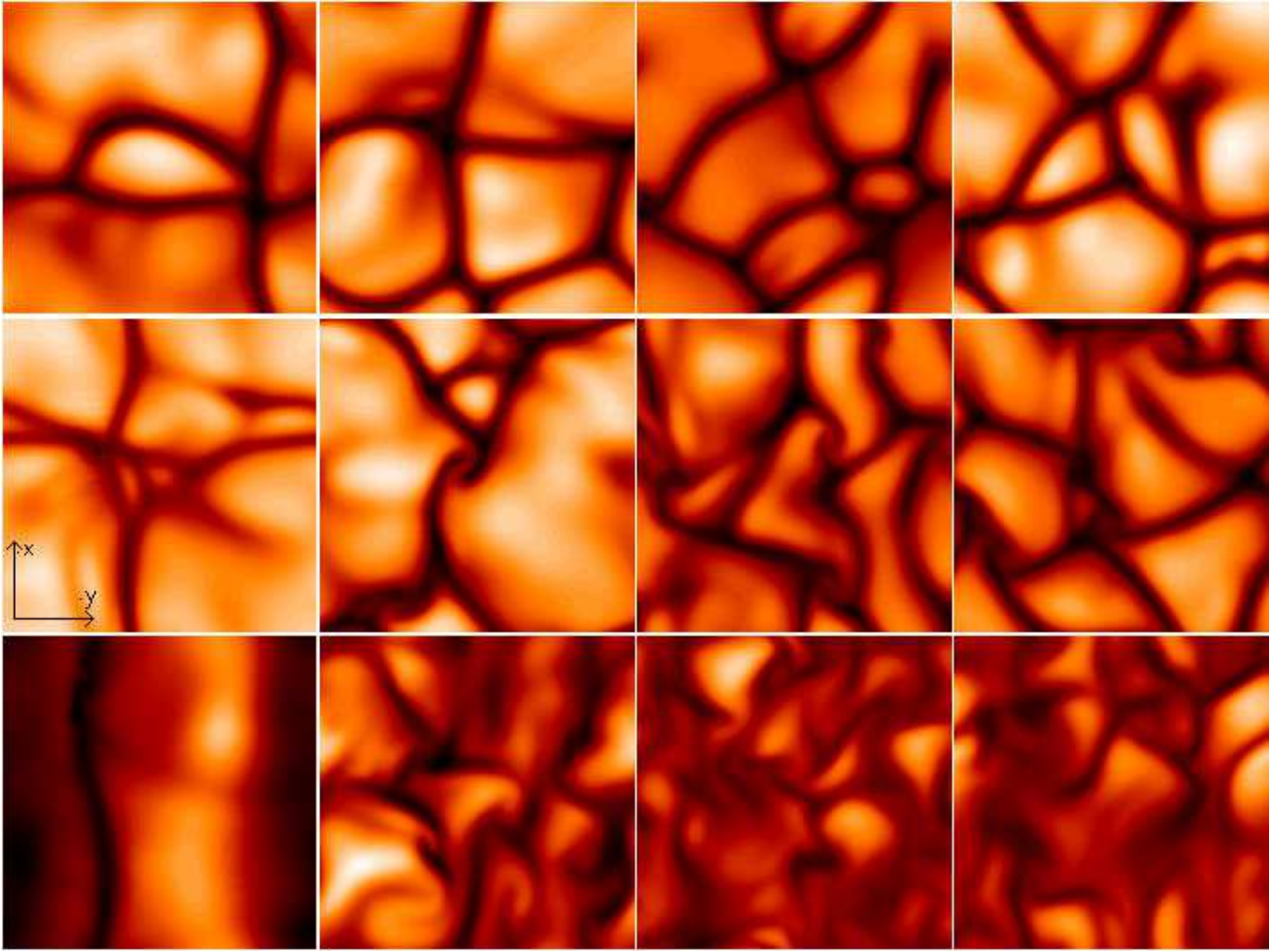}
      \caption{Horizontal cross-sections of temperature from fully
      developed convection at the top of the convectively unstable
      layer, $z = z_1$. From left to right, the individual frames are
      from latitudes $0$, $-30$, $-60$, and $-82.5$ degrees. The
      rotation rate increases from top to bottom, corresponding to
      sets Co01, Co1, and Co10. Here, the $x$-axis points to the east
      and the $y$-axis to the north.}
      \label{pic:eet}
   \end{figure*}

   \begin{figure}
   \centering
      \includegraphics[width=0.5\textwidth]{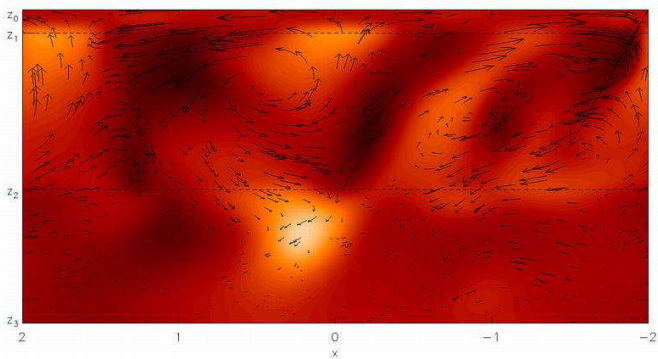}
      \caption{Vertical cross-section of temperature fluctuations and
      meridional velocity vectors in the $xz$-plane at $y = -1$ in the
      run Co7-60. The arrow at the upper right corner indicates the
      direction of the rotation vector.}
      \label{pic:slice_xz}
   \end{figure}

   \subsection{Convective structures} 
   \label{subsec:convstru}

   To illustrate the general appearance of the structures arising in
   the present calculations we show a typical snapshot of temperature
   fluctuations overplotted with the velocity vectors from fully
   developed convection in a moderate rotation case HICo1-30 in
   Fig.~\ref{pic:boxplot}. As in many previous studies we find a
   convection pattern dominated by broad warm upflows and narrow cool
   downflow plumes. Near the top the upflows form cells, separated
   from each other by a network of downflow structures at the cell
   boundaries. The downflow structures remain coherent over large
   depths, sometimes even extending over the whole convectively
   unstable layer. Generally, the horizontal scale of the convective
   pattern is observed to decrease as a function of depth. Note that
   the downflow lanes seen near the top tend to form a few
   well-defined plumes at larger depths and the regions of upward flow
   connect and become broader. However, the upflows also tend to
   become less well-defined, so that much more small-scale structure
   is seen as opposed to the large and well-defined 'granules' near
   the top. Although this apparently contradicts the prediction of the
   mixing length theory, on average the upflows occupy a larger area
   in the deeper layers in accordance with the basic mixing length
   concept.

   The effects of rotation can be studied from Fig.~\ref{pic:eet} where
   we show horizontal contours of temperature from three different
   calculation sets at four latitudes. In the weak rotation case (set
   Co01), shown in the uppermost row, the convective cells have more or
   less angular shapes, which pattern changes only slightly as function
   of latitude. There is a weak tendency of the sizes of the structures
   to get smaller towards the pole, which effect becomes more pronounced
   as the rotation is increased (the middle row, set Co1); the size of
   the structures clearly decreases as function of increasing latitude.
   The angular appearance of the cells also changes due to the strong
   vortical downflows which are generated by the interaction of the
   converging flow with the Coriolis force at the cell edges. In the
   rapid rotation case (the bottom row, set Co10) the convective pattern
   is dominated by rather irregular small-scale structures, except for
   the equatorial case, where the pattern is totally aligned with the
   rotation axis, reminiscent of the banana cells seen in some global
   convection models (e.g. Brun \& Toomre \cite{BruTo02}). Such an
   alignment is a generic feature of the convective motions in the rapid
   rotation regime, as can be seen from Fig.~\ref{pic:slice_xz}, where
   $xz$-slices of temperature fluctuations overplotted with velocity
   field vectors are shown from the run Co7-60.

   \subsection{Velocity field characteristics}
   \label{subsec:velcha}	

   In Tables \ref{tab:Calcu} and \ref{tab:HiresCalcu} we have calculated
   some diagnostics of the velocity field. Comparing the total and
   fluctuating rms-velocities (columns 6 and 7, respectively), one can
   note that both tend to decrease towards the poles and increase
   steeply near the equator for rapid rotation. A similar trend can be
   seen in the azimuthal and vertical velocity field components (columns
   9 and 10). This is partly due to the fact that the convection is more
   efficient in the equatorial regions compared to the polar ones (see
   Sect.~\ref{subsec:conveff}), partly because the increasing rotational
   influence generates azimuthal mean flows, seen as a deviation between
   the total and the fluctuation, which are strongest near the equator.

   The degree of horizontal anisotropy, measured by the quantity
   \begin{eqnarray}
      A_{H} = \frac{\langle u_{y}'^{2} \rangle - \langle u_{x}'^{2}\rangle}{u_{t}^{2}}\;. \label{equ:AH}
   \end{eqnarray}
   is very small at all latitudes for the slow rotation cases (sets
   Co01 to Co1). As rotation is increased, $A_H$ obtains positive
   values near the equator due to the strong azimuthal mean flows
   generated there. However, the isotropy is more or less
   re-established at southern latitudes $\Theta > 30\degr$, also for
   the rapid rotation cases.

   The vertical component of the velocity is dominating over the
   horizontal ones for the slow rotation cases, which means that the
   preferred direction of the convective motions is the radial one, in
   which stratification and energy transport also occur. This
   anisotropy, measured by the quantity
   \begin{eqnarray}
      A_{V} = \frac{\langle u_{y}'^{2} \rangle - \langle u_{z}'^{2}\rangle}{u_{t}^{2}}\;, \label{equ:AV}
   \end{eqnarray}
   is observed to decrease as the rotational influence increases,
   because the convective structures become more and more aligned with
   the rotation vector. The sign of this quantity is also useful since
   it should coincide with the sign of the radial gradient of the
   angular velocity, $\frac{d \Omega}{d r}$ (R\"udiger
   \cite{Rudiger80}). In the present calculations, $A_{V}$ is almost
   always negative, except for the rapid rotation cases near the
   equator. Thus, the velocity anisotropy generated in our calculations
   reproduces the helioseismic result of positive $\frac{d \Omega}{d r}$
   at the equator and at low latitudes and a negative one at higher
   latitudes only for the sets Co4 to Co10. However, we expect that this
   result is slightly affected by the fact that the averages were taken
   only over the convectively unstable region. In this case the flow
   regions where the vertical flows turn into horizontally diverging
   flows, mostly occurring in the upper boundary and overshoot layers,
   were excluded from the calculation, resulting in the overestimation
   of $u_z$ over $u_y$.

   \begin{figure}[t]
   \centering
      \includegraphics[width=0.5\textwidth]{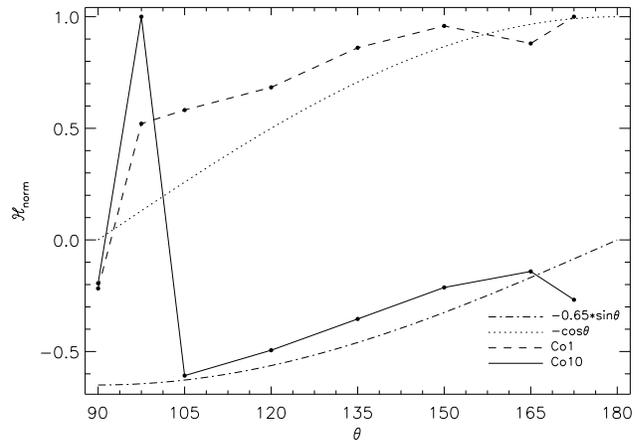}
      \caption{Volume-averaged kinetic helicity, $\mathcal{H}$,
      normalised by the maximum value, for the sets Co1 (dashed line)
      and Co10 (solid line). For comparison, the dotted line shows a
      $-\cos \theta$ profile and the dash-dotted line a $-\sin \theta$
      profile multiplied by 0.65 in order to fit to the data.}
      \label{pic:heli}
   \end{figure}

   \begin{figure}
   \centering
   \includegraphics[width=0.5\textwidth]{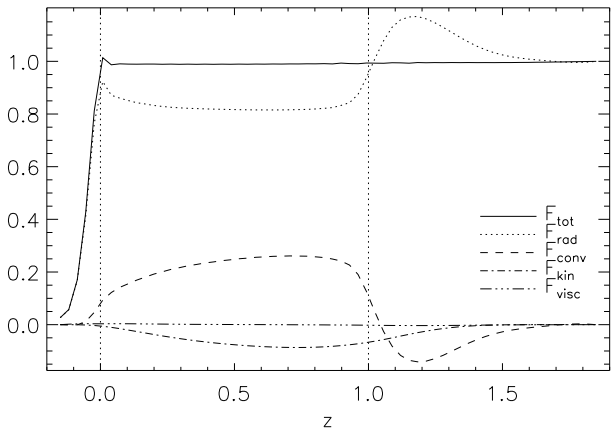}
      \caption{Mean energy fluxes as functions of depth from the run
      Co1-30, normalised by the input flux, see
      Sect.~\ref{subsec:conveff} for definitions.}
      \label{pic:flux}
   \end{figure}

   \begin{figure*}[t]
   \centering
      \includegraphics[width=1.\textwidth]{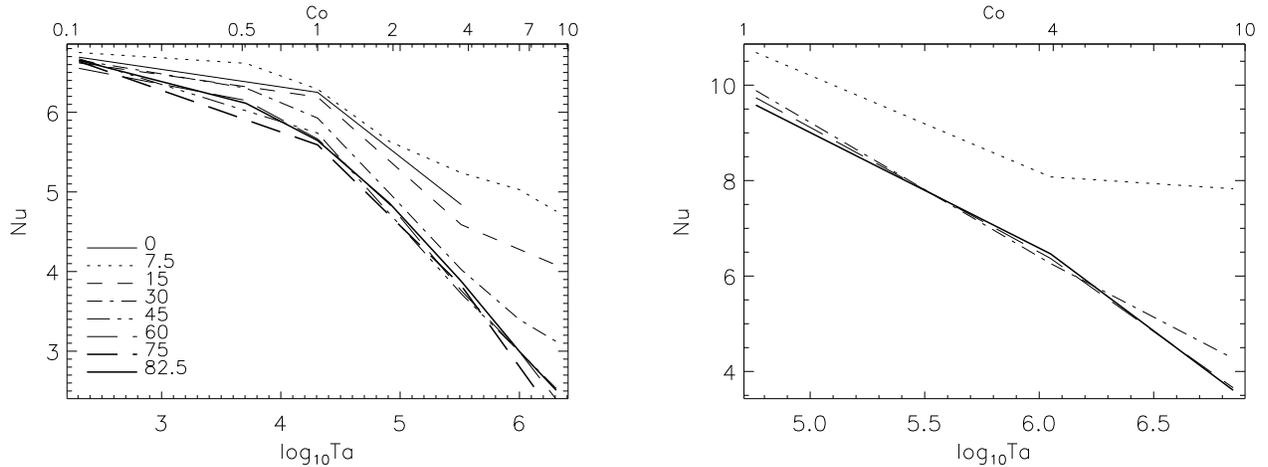}
      \vspace{-7cm}
      \caption{The Nusselt number as a function of Taylor (lower
      x-axis) and Coriolis (upper x-axis) numbers at different
      latitudes for the $\rm Ra = 2.5 \cdot 10^{5}$ (left) and $\rm Ra
      = 10^{6}$ (right) calculations. Different linestyles are used
      for each latitude investigated, as indicated in the left panel.}
      \label{pic:nusselt}
   \end{figure*}

   \subsection{Kinetic helicity}
   \label{subsec:heli}
   Column 11 in Tables \ref{tab:Calcu} and \ref{tab:HiresCalcu} shows
   the volume averaged kinetic helicity, $\mathcal{H}$, defined by
   \begin{eqnarray} \mathcal{H} = \langle (\nabla \times \vec{u}') \cdot
   \vec{u}' \rangle\;.  \end{eqnarray} This quantity is interesting in
   that it is believed to play a major role in the dynamo theory,
   providing the so-called $\alpha$-effect, which describes the
   generation of large-scale poloidal magnetic fields due to the
   inductive action of helical turbulence. In the simplest
   approximation, making use of the first order smoothing and
   assumption of isotropy, $\alpha \approx - \frac{1}{3} \tau'
   \mathcal{H}$, where $\tau'$ is the correlation time associated with
   the turbulence. It has been noted that the $\alpha$ computed from
   magnetoconvection calculations qualitatively agrees with the above
   approximation (Ossendrijver et al. \cite{Osse01},
   \cite{Osse02}). Furthermore, it is possible to determine the sign
   and latitudinal distribution of the magnitude of $\mathcal{H}$ in
   the surface layers of the Sun from helioseismology (Duvall \& Gizon
   \cite{DuGi00}; see also R\"udiger et al. \cite{Rudetal99} who
   studied numerically the determination of $\mathcal{H}$ from surface
   flows). The aforementioned studies indicate that $\mathcal{H}$ is
   maximal at the poles with a nearly $\cos \theta$-latitude
   dependence. The sign of $\mathcal{H}$ is noted to be positive in
   the bulk of the convection zone in the southern hemisphere and
   negative in the northern hemisphere.

   We find that on the low rotation calculations $\mathcal{H}$ shows
   several sign changes as function of depth so that the volume
   average is small and does not show a clear latitudinal trend. When
   rotation is moderate (sets Co1 to Co4) there is a clear tendency
   for the helicity to be positive in the convectively unstable region
   and to increase towards the pole (bar the Co4 set with Ra $2.5
   \cdot 10^{5}$), in accordance with the aforementioned studies. As
   function of depth, the positive values peak near the top of the
   convection zone while a region of smaller negative helicity resides
   in the overshoot layer. This region of negative $\mathcal{H}$ grows
   as rotation is increased so that for the strong rotation cases,
   only a narrow region of positive helicity remains in the surface
   layers. Near the equator, this positive peak dominates in the
   volume average of $\mathcal{H}$, but for latitudes $\Theta <
   -15\degr$ the sign is negative which behaviour carriers all the way
   to the pole, however, with monotonically decreasing absolute values
   (see Figure \ref{pic:heli}). Taken that the close relation between
   the helicity and the $\alpha$-effect persists in the rapid rotation
   regime, this would indicate that for strong enough rotation, the
   sign of the $\alpha$-effect changes from negative to positive and
   simultaneously decreases in magnitude at higher latitudes, whilst
   larger negative values occur in the equatorial regime. Interpreted
   in terms of the classical dynamo theory, in the regime of slow and
   moderate rotation, the poloidal field generation would be more
   efficient near the poles with equatorward migration of activity
   phenomena, when combined with the observed internal rotation
   law. In the rapid rotation regime, in contrast, the $\alpha$-effect
   would be more efficient in the equatorial regions, and the
   migration would be equatorward at low latitude regions and poleward
   at higher latitudes. However, the proportionality between
   $\mathcal{H}$ and the $\alpha$-effect is only approximate, and can
   yield incorrect results for realistic circumstances. Thus,
   calculations with magnetic fields are needed in order to determine
   whether the behaviour suggested by the helicity really occurs in
   the rapid rotation regime.

   \subsection{Convection efficiency}
   \label{subsec:conveff}	
   To show the energy transport in a typical calculation, we plot the
   depth dependence of the horizontally averaged energy fluxes in
   Fig.~\ref{pic:flux} for the run Co1-30. We define
   \begin{eqnarray}
      F_{\rm rad} & = & \kappa \frac{\partial \langle e \rangle}{\partial z}\;, \\
      F_{\rm conv}& = & -\gamma \langle \rho e' u_z' \rangle\;, \\
      F_{\rm kin} & = & -\frac{1}{2} \langle \rho \vec{u}^{2} u_{z} \rangle\;, \\
      F_{\rm visc}& = & - 2 \nu \langle \rho S_{zi} u_i \rangle\;,
   \end{eqnarray}
   as the radiative, convective, kinetic, and viscous fluxes. The
   minus signs arise because we define positive flux to be directed
   outwards. Most of the energy is transported by radiative diffusion,
   whilst the convection transports roughly one fifth of the total
   energy flux through the convection zone. The flux of kinetic energy
   is directed downwards, and amounts to a few per cent of the total
   flux. The viscous flux is in all cases negligible. The change of
   sign of $F_{\rm conv}$ in the overshoot layer is due to the fact
   the negative temperature fluctuations entering the stable layer
   find themselves as positive fluctuations in the new background
   stratification. We find that the convective flux decreases as
   rotation is increased so that $F_{\rm conv}$ in the Co10 set is
   roughly half of that in the Co01 set.

   The rotational influence on the convection efficiency can be
   estimated quantitatively by the Nusselt number, (see
   e.g. Brandenburg et al. \cite{Brand90})
   \begin{eqnarray}
     {\rm Nu} = \frac{F_{\rm tot} - F_{\rm ad}}{F_{\rm rad}^{(0)} - F_{\rm ad}}\;, \label{equ:nusselt}
   \end{eqnarray}
   where $F_{\rm tot} = F_{\rm rad} + F_{\rm conv} + F_{\rm kin} +
   F_{\rm visc}$ is the total flux through the convectively unstable
   layer, $F_{\rm ad} = \kappa \big( \frac{\partial e}{\partial z}
   \big)_{\rm ad}$ the radiative flux if the layer is adiabatic, and
   $F_{\rm rad}^{(0)} = \kappa \frac{\Delta e}{\Delta z}$ the
   radiative flux if the temperature stratification in the unstable
   layer is linear. The Nusselt number gives the ratio of the total
   heat flux and the heat flux which would be transported by
   conduction alone. Thus Nu must exceed unity for convection to be
   present. 
  
   The resulting Nusselt numbers as functions of rotation, measured by
   Ta and Co, at the latitudes investigated are shown in
   Fig.~\ref{pic:nusselt}. The slow rotation case Co01 does not show
   any definite latitudinal variation, which is indeed the expected
   behaviour. The variation seen within the set Co01 can then be used
   as an error estimate for the other sets. When the rotation is
   stronger, the efficiency of convection decreases as a function of
   latitude in a more or less monotonic fashion. Convection at
   equatorial regions seems to be most vigorous and for $\Theta \le
   -45\degr$ the efficiency does not change appreciably as the
   latitude increases. The trend that convection at polar regions
   would be more vigorous than at some intermediate latitudes noted by
   Gilman (\cite{Gilman77}) and Pulkkinen et al. (\cite{Pulkki93}) is
   not observed here. Furthermore, we do not find a regime where Nu
   increases with increasing Ta (Hathaway \& Somerville
   \cite{Hatha83}). For the $\rm Ra = 10^{6}$ calculations the
   qualitative results are very similar, the distinction being the
   roughly 50\% larger values of Nu.

   \subsection{Overshooting depth}
   \label{subsec:overshoot}
   Convective overshooting is thought to be important for example in
   the context of stellar magnetism providing means to store magnetic
   flux below the convection zone by resisting the buoyancy effect. In
   our calculations, indeed, the downflows very often overshoot into
   the stable layer below, where they decelerate due to the buoyancy
   force. We define the depth of overshooting as
   \begin{eqnarray}
   d_{\rm os} = z_{2} - z_{\rm os},
   \end{eqnarray}
   where $z_{\rm os}$ is the depth where the horizontally averaged
   kinetic energy flux falls below a threshold value of $e^{-1}$ of its
   value at the bottom of the convectively unstable layer. In order to
   facilitate a comparison to a real star we normalise the overshooting
   depth by the pressure scale height, $H_{\rm p2} = e_{2}(\gamma -
   1)/g$, at the bottom of the convectively unstable layer.

\begin{figure}
   \centering
   \includegraphics[width=0.5\textwidth]{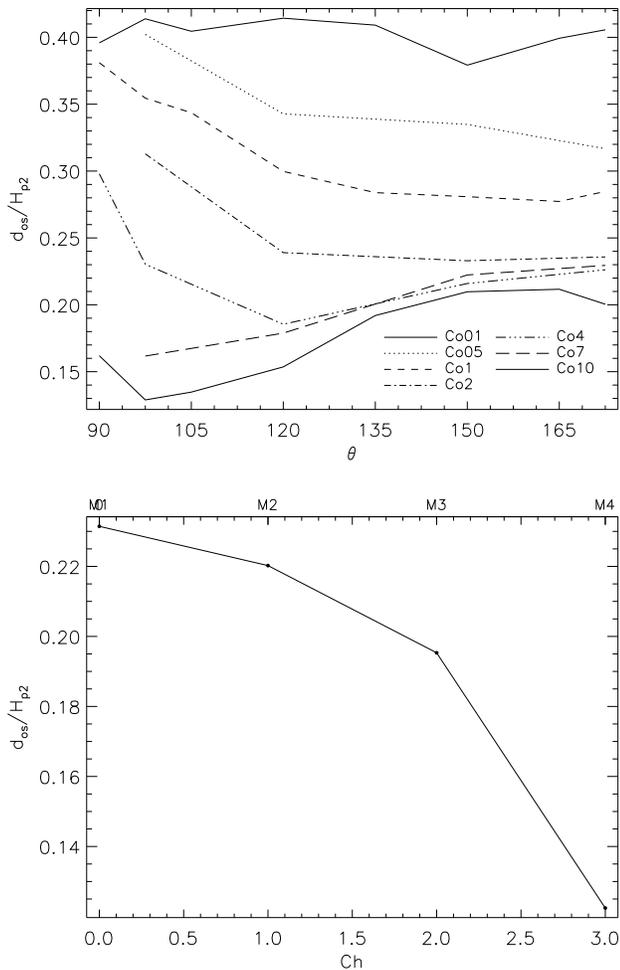}
      \caption{Upper panel: overshooting depth, $d_{\rm os}$, as
       function of latitude for the calculations with Ra = $2.5 \cdot
       10^{5}$.  Different linestyles are used for each set of
       calculations, differing by Coriolis number, as indicated in the
       figure. Lower panel: $d_{\rm os}$, as function of the
       Chandrasekhar number. The upper $x$-axis denotes the
       corresponding runs presented in Table~\ref{tab:magruns}.}
       \label{pic:overshoot}
   \end{figure}

   The calculated depths for the Ra $=2.5 \cdot 10^5$ runs are shown
   in the upper panel of Fig.~\ref{pic:overshoot} as functions of
   latitude and Coriolis number. As a general feature we find that the
   overshooting decreases as a function of increasing rotation at a
   given latitude. Such behaviour has earlier been reported by
   Brummell et al. (\cite{Brumm02}) and Ziegler \& R\"udiger
   (\cite{ZieRu03}). For the slowest rotation investigated (set Co01)
   the overshooting extends rather deep, down to $0.4 H_{\rm p2}$, and
   the latitude dependence is very weak as expected. For moderate
   rotation (sets Co05 to Co2) there is a clear trend of decreasing
   overshoot depth with latitude. However, as rotation is increased
   further, there is an opposite trend. At the poles, $d_{\rm os}$
   seems to stay more or less constant, whilst it continues to
   decrease at the equator. This may be interpreted to be due to the
   stronger influence of the Coriolis force on the vertical downflows
   at the equatorial regions, where they tend to align with the
   rotation vector. The lower panel of Fig.~\ref{pic:overshoot} shows
   $d_{\rm os}$ as function of the Chandrasekhar number for the
   magnetic runs (see Sect. \ref{subsubsec:maglam} for a detailed
   description of the runs). We find that the overshooting depth is
   reduced monotonically with increasing imposed field strength for
   the present range of parameters where the ratio of the gas to
   magnetic pressure in the bottom of the convection zone spans from
   $\approx 2.6\cdot 10^5$ to 260. Increase of $d_{\rm os}$ for small
   magnetic field strengths, noted by Ziegler \& R\"udiger
   (\cite{ZieRu03}), is not observed.

   For the most rapid rotation, $d_{\rm os}$ varies between about $0.1
   H_{\rm p2}$ at the equator to about $0.2 H_{\rm p2}$ at high
   latitudes. This value is still larger than the estimates from
   helioseismology (e.g. Monteiro et al. \cite{Monte94}), but the
   rotational influence in our study is most probably still weaker
   than in the deeper layers of the solar convection zone. One must,
   however, bear in mind that the normalised energy flux, $F_{\rm
   tot}/\rho c_{\rm s}^{3}$, where $c_{\rm s}$ is the sound speed, at
   the bottom of the convectively unstable layer is roughly eight
   orders of magnitude larger than in the case of the
   sun. Furthermore, for non-rotating convection, the penetration
   depth scales with $u_{z}^{3/2}$ and the overshooting depth with the
   square root of the input flux (e.g. Zahn \cite{Zahn91}; Hurlburt et
   al. \cite{Hurlburt94}; Singh et al. \cite{Singh98}). The difference
   between penetration and overshooting is that the former extends the
   superadiabatic region and the latter does not. In the present
   calculations the thermal stratification does not appreciably change
   due to the overshooting and thus the latter scaling relation should
   be applicable. However, in our results no such scaling exists,
   since the input flux in the Ra=$10^{6}$ calculations is half the
   one used in the Ra=$2.5\cdot10^{5}$ case but the overshooting depth
   decreases only by a factor of ten to fifteen per cent for the Co1
   and Co4 sets and actually increases for the Co10 set (not shown). A
   possible explanation of this behaviour is that the mixing length
   theory, assumptions of which were used in the derivation of the
   scaling relation, breaks down when rotation is present. On the
   other hand, it has to be noted that our description of the
   overshoot region is not very realistic, the thermal diffusivity
   profile being essentially a smoothed step function, which quite
   probably acts to hinder the efficiency of overshooting in
   comparison to a smoothly varying profile.

   \subsection{Mean flows}
   \label{subsec:meanflows}

   \begin{figure*}
   \centering
   \includegraphics[width=0.9\textwidth]{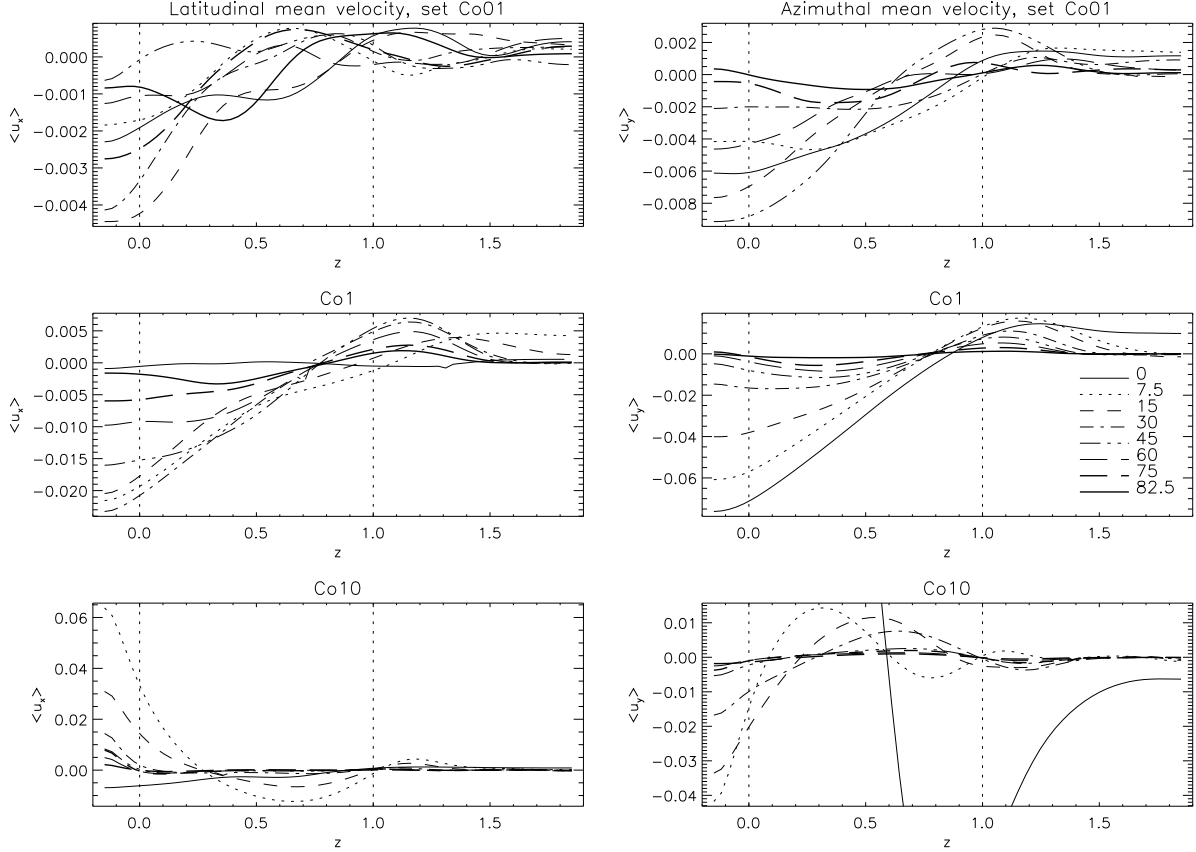}
      \caption{Latitudinal (left) and azimuthal (right) mean flows,
      $\langle u_{x} \rangle$ and $\langle u_{y} \rangle$,
      respectively, plotted against depth from sets Co01 (top, slow
      rotation), Co1 (middle, moderate rotation), and Co10 (bottom,
      rapid rotation). The different linestyles correspond to
      different latitudes, which are denoted in the middle figure. On
      the lowest graph on the right, only a part of the equatorial
      azimuthal velocity is shown due to its large magnitude.}
       \label{pic:meanu}
   \end{figure*}

   \begin{figure*}
   \centering
   \includegraphics[width=1.0\textwidth]{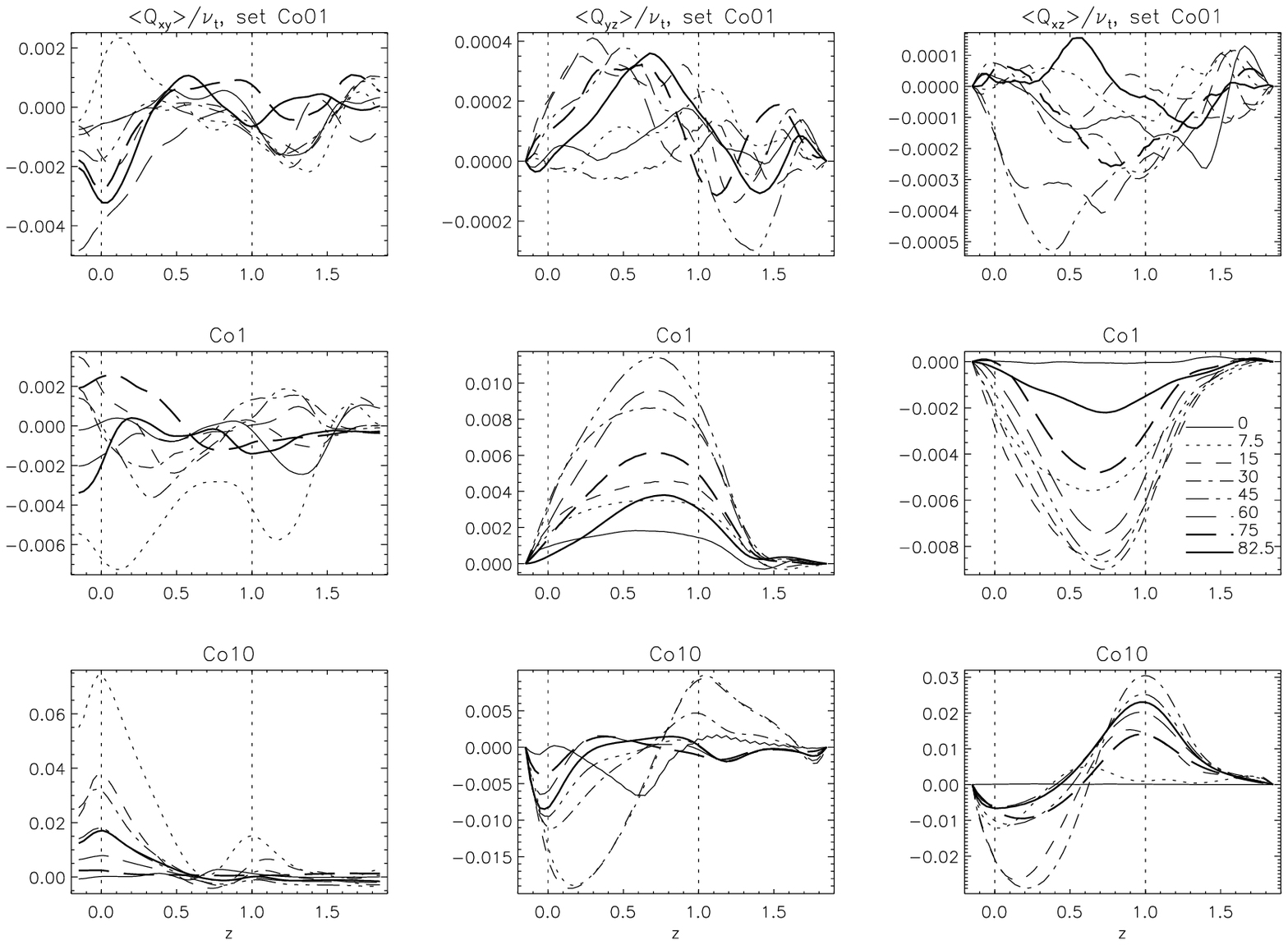}
      \caption{Reynolds stress components $\langle Q_{xy}
      \rangle/\nu_{t}$, $\langle Q_{yz} \rangle/\nu_{t}$, and
      $\langle Q_{xz} \rangle/\nu_{t}$ as functions of depth for
      the sets Co01 (top), Co1 (middle), and Co10 (bottom),
      respectively. Different latitudes are represented by the
      linestyles denoted in the middle figure.}
       \label{pic:rey_hori}
   \end{figure*}

   The generation of differential rotation by the presence of
   anisotropic turbulence and uniform rotation is accounted for in the
   theory of the $\Lambda$-effect (e.g. R89) by the interaction of the
   anisotropic eddies with the Coriolis force. In the rectangular
   domain used in this study, this effect works to generate mean
   flows, of which of particular interest are the latitudinal and
   azimuthal ones which are indicative of the differential rotation in
   a full spherical domain. However, one must bear in mind that the
   boundary conditions in a spherical domain would permit a variety of
   other effects which affect the mean flows, such as horizontal
   temperature gradients, which in the present geometry could only be
   imposed and cannot arise naturally.

   To investigate the depth dependence of the mean flows we average the
   horizontal velocities over a time period designated in Table
   \ref{tab:Calcu} and over the horizontal directions. This procedure
   yields $\langle u_{x} \rangle(z)$ and $\langle u_{y} \rangle(z)$ for
   the latitudinal and azimuthal mean velocites, respectively, which are
   functions of $z$ alone. The vertical mean velocity $\langle u_z
   \rangle(z)$, calculated in this way, is very small in comparison to
   the horizontal ones due to the boundary conditions which do not
   permit a net vertical momentum flux. The fact that $\langle u_z
   \rangle$ still has a small finite value is due to the different
   filling factors of up- and downflows. Fig.~\ref{pic:meanu} shows
   typical examples of these flows plotted at different latitudes for
   slow (Co01), moderate (Co1), and rapid rotation (Co10).

   The predominant feature of the latitudinal flow for slow and
   intermediate rotation is the mostly negative
   (poleward\footnote{Note that the box lies on the southern
   hemisphere}) flow in the bulk of the convection zone and the
   positive (equatorward) flow in the overshoot layer. This trend is
   strongest for the set Co2, and when rotation increases, the
   latitudinal flow starts to diminish and the direction reverses for
   the most rapid rotation. For the rapid rotation cases (Co7 and
   Co10) there is a strong equatorward flow confined to a narrow layer
   near the surface and a much weaker poleward flow in the bulk of the
   convection zone. In the overshoot layer only a very weak
   equatorward flow is present.

   The mean latitudinal flows in the slow rotation regime agree
   with the results of Chan (\cite{Chan01}) and Brummell et
   al. (\cite{Brumm98}). When rotation is stronger, the qualitative
   character of Chan's results stays the same, in disagreement with
   ours, whereas in the study of Brummell et al. (\cite{Brumm98}) a
   sign change near the top of the domain is reported (their
   Fig.~6). The increasing magnitude of the latitudinal flow towards
   the equator noted by Brummell et al. (\cite{Brumm98}) is also
   present in our results for moderate rotation. However, the
   magnitude of the latitudinal flow in our study seems to grow for
   increasing rotation, a feature not present in the two
   aforementioned studies.

   The azimuthal mean flow shows generally very similar behaviour,
   with increasingly negative values (westward motion) near the
   surface and smaller positive values (eastward) in the overshoot
   layer for slow and moderate rotation (up to the set Co2). The
   magnitude of the azimuthal flow decreases more or less linearly as
   function of depth, and also as function of increasing latitude. The
   latter effect is reminiscent of the Taylor-Proudman theorem
   (e.g. Chandrasekhar \cite{Chandra61}) which states that motions
   cannot vary in the direction of the rotation axis leading to
   cylinder shaped contours of the angular velocity. The former effect
   was discussed by Chan (\cite{Chan01}), who showed that the dominant
   term in the equation for the mean azimuthal flow at the equator is
   the horizontal component of the Coriolis force, $f_{\rm C} =
   2\,\Omega \cos \Theta u_{z}$, which produces a shear of approximate
   magnitude of
   \begin{eqnarray} 
     \frac{\Delta \langle u_{y} \rangle}{\Delta z} \approx 2\,\Omega\;, \label{equ:duyChan}
   \end{eqnarray} 
   where the difference $\Delta$ is taken over the
   unstable layer. Our results follow this relation for the slow and
   moderate rotation cases (sets Co01 to Co2), but this relation fails
   for the rapid rotation regime. However, a more general way of
   explaining the mean flows in the present calculations can be easily
   obtained by considering the Reynolds stresses, see
   Sect. \ref{sec:mere} below.

   Just as the latitudinal flow, also the azimuthal mean flow reverses
   direction in the bulk of the convection zone for the most rapid
   rotation (Co7 and Co10). There is also a moderate westward flow in
   the bulk of the convection zone and a layer of negative shear near
   the surface. A narrow layer of retrograde motion exists in the
   overshoot region. This behaviour is very similar to the
   investigations by Brummell et al. (\cite{Brumm98}) and Chan
   (\cite{Chan01}).

   \subsubsection{Generation of the mean flows by the Reynolds stresses}
   \label{sec:mere}

   Next we study the horizontal momentum balance in the model by
   deriving the equations for the latitudinal, azimuthal, and vertical
   momentum balance, which are obtained by taking a horizontal average of
   the momentum equation, Eq.~(\ref{equ:momentum})
   \begin{eqnarray}
     -\frac{\partial}{\partial z} (\langle \rho u_{x} u_{z} \rangle) - 2\,\Omega \langle \rho u_{y} \rangle \sin \Theta & = & 0\;, \label{equ:momx}\\ 
     -\frac{\partial}{\partial z} (\langle \rho u_{y} u_{z} \rangle) + 2\,\Omega \langle \rho u_{x} \rangle \sin \Theta & = & 0\;, \label{equ:momy}\\
     -\frac{\partial}{\partial z} (\langle \rho u^{2}_{z} \rangle) -
2\,\Omega \langle \rho u_{y} \rangle \cos \Theta - \frac{\partial
\langle p \rangle}{\partial z} + \langle \rho \rangle g & = & 0\;. \label{equ:momz}
   \end{eqnarray}
   In the azimuthal equation we have dropped the Coriolis term
   involving the mean vertical momentum $\langle \rho u_{z} \rangle$,
   which vanishes due to the boundary conditions. Furthermore,
   applying the Reynolds rules, the remaining terms of the form
   $\langle u_{i} u_{j} \rangle$ can be expanded using
   Eq.~(\ref{equ:corr}). In this expansion terms with $\langle u_{z}
   \rangle$ appear, but these are very small in comparison to the
   other mean velocities and can be neglected. The remaining terms of
   Eqs. (\ref{equ:momx}) to (\ref{equ:momz}) can then be used to solve
   for the mean horizontal velocities
   \begin{eqnarray} 
   \langle u_{y} \rangle & = & - \frac{1}{2\,\Omega \langle \rho \rangle \sin
     \Theta} \frac{\partial}{\partial z} (\langle \rho \rangle \langle
     u'_{x} u'_{z} \rangle)\;, \label{equ:meanuy1}\\ 
   \langle u_{x} \rangle & = & \frac{1}{2\,\Omega \langle \rho \rangle \sin 
     \Theta} \frac{\partial}{\partial z} (\langle \rho \rangle \langle u'_{y}
     u'_{z} \rangle)\;,\\ 
   \langle u_{y} \rangle & = & - \frac{1}{2\,\Omega \langle \rho \rangle \cos
     \Theta} \Big[ \frac{\partial}{\partial z} (\langle \rho \rangle \langle
     u_{z}'^{2} \rangle) + \frac{\partial \langle p \rangle}{\partial z}
     - \langle \rho \rangle g \Big]\;.\label{equ:meanuy2} 
   \end{eqnarray}
   At the equator, the latitudinal mean flow vanishes due to symmetry,
   and Eq.~(\ref{equ:meanuy1}) has a singularity. However,
   Eq.~(\ref{equ:meanuy2}), which includes contribution of the
   diagonal Reynolds stress component $\langle u_{z}'^{2} \rangle$ in
   the pressure field, can be used instead. We find that the vertical
   derivatives of the Reynolds stress components $\langle u'_{y}
   u'_{z} \rangle$ and $\langle u'_{x} u'_{z} \rangle$ explain the
   generated horizontal mean flows in all non-equatorial models. At
   the equator, Eq.~(\ref{equ:meanuy2}) reproduces the observed zonal
   mean flow, and is applicable at all latitudes investigated. Thus,
   the observed mean flows in the models arise due to the
   non-vanishing transport of momentum driven by the Reynolds stresses
   as expected.
   
   \subsection{Reynolds stresses and the $\Lambda$-effect}
   \label{subsec:reynolds} In this section we discuss the spatial
   distribution of the Reynolds stress components generated in the
   calculations, and their dependence on rotation and large-scale
   magnetic fields. To draw conclusions of the effects of the generated
   stresses on the angular momentum transport and generation of
   differential rotation in a global spherical geometry, we parametrise
   the stresses in terms of the $\Lambda$-effect coefficients. This is
   also interesting because the calculations presented here are not
   restricted by the assumptions of FOSA, due to which the results may
   be used to test the validity of this approximation. It has to be born
   in mind, however, that our local calculations also have restrictions,
   including the effects of periodic horizontal boundaries, limited
   number of pressure and density scale heights included, and the small
   Reynolds and Rayleigh numbers, due to which the turbulence modelled
   is relatively weak and the convection rather inefficient.

   \subsubsection{The stresses due to the generated mean flows} As
   discussed in the previous section, mean flows exhibiting radial
   gradients naturally arise in the calculations due to the non-zero
   Reynolds stresses (they also vary as a function of latitude
   measured from different calculation sets, but due to the periodic
   horizontal boundaries the latitudinal gradients vanish in the
   calculations). In this case, in the series expansion of the
   Reynolds stresses, Eq.~(\ref{equ:qlamdif}), the diffusive
   contribution will become non-zero as well, due to which it should
   be separated before calculating the $\Lambda$-coefficients, for
   which we have adopted the following procedure. Using the radial
   profiles of the latitudinal and azimuthal mean flows generated in
   the calculation of interest as an additional imposed velocity field
   component stationary in time (i.e. writing the velocity field as
   ($u_x + u_x^{\rm shear}(z)$,$u_y + u_y^{\rm shear}(z)$,$u_z$) and
   adding advection terms of types $u_i^{\rm shear} \partial
   u_j/\partial{x_i}$ and $u_z \partial u_{i}^{\rm shear}/\partial{z}$
   to the equations), we repeat a few calculations switching off
   rotation, and calculate the generated Reynolds stresses. These
   stresses now correspond to the diffusive part in the series
   expansion, denoted here by $Q_{ij}^{(\rm shear)}$. Subtracting
   these stresses from the ones generated in the full calculations, we
   obtain the non-diffusive contribution
   \begin{eqnarray} Q_{ij}^{(\Lambda)} = Q_{ij}^{(\rm tot)} -
   Q_{ij}^{(\rm shear)}\;.  \end{eqnarray} We chose to apply this
   procedure only to a subset of the full calculations, namely for the
   sets Co1 and Co10, corresponding to moderate and rapid rotation,
   respectively, and at the latitudes of $0\degr$, $-7.5\degr$,
   $-30\degr$, and $-60\degr$; the results are summarised in
   Table~\ref{tab:shearstress}.

   We note that the stresses generated by the shear flows are largest
   in the cases where the imposed flows were taken from the equatorial
   or near-equatorial cases. The stress component $Q_{xy}^{(\rm
   shear)}$ is small in comparison to $Q_{xy}^{(\rm tot)}$ for all
   latitudes in the Co1 set (note that the stress components $Q_{xy}$
   and $Q_{xz}$ should vanish at the equator due to symmetry). In the
   rapid rotation case $Q_{xy}^{(\rm shear)}$ is approximately one
   fifth of the total stress near the equator and about one eight for
   latitude $-30 \degr$. For the other two latitudes the stresses are
   statistically consistent with zero. Since the contribution of the
   shear is of the same sign as the total, we note that it acts as to
   diminish the contribution of the $\Lambda$-effect. We thus conclude
   that the horizontal $\Lambda$-effect derived from the total stress
   is overestimated by maximally 20 per cent near the equator for the
   most rapid rotation cases.

   The stress component $Q_{yz}$ shows an opposite trend:
   $Q_{yz}^{(\rm shear)}$ is in general of opposite sign to
   $Q_{yz}^{(\rm tot)}$ which indicates an underestimation of the
   vertical $\Lambda$-effect. This underestimation is largest near the
   equator due to the large azimuthal mean flows there (see Figure
   \ref{pic:meanu}). Noteworthy is that the shear contribution is
   almost equal to the $\Lambda$ part at the equator for the Co10 case
   but for the latitude $-7.5 \degr$ the shear contribution is already
   insignificant. In conclusion, the vertical $\Lambda$-effect
   estimated from $Q_{yz}^{(\rm tot)}$ is in general underestimated
   near the equator possibly by a factor of two to three but that this
   effect is significantly weaker for all higher latitudes, being
   maximally a few tens of per cents in the moderate rotation cases.
   The stress component $Q_{xz}^{(\rm shear)}$ is insignificantly
   small in comparison to $Q_{yz}^{(\rm tot)}$ for all non-equatorial
   cases studied.

   \begin{table}
   \centering
      \caption[]{Comparison of the volume averaged Reynolds stresses
      generated by rotation and dynamically generated mean flows
      versus stresses generated by mean flows alone (in units of
      [$10^{-3}]$).}
      \vspace{-0.75cm}
      \label{tab:shearstress}
     $$
         \begin{array}{p{0.24\linewidth}cccc}
            \hline
            \noalign{\smallskip}
            Run      & \langle Q_{xy} \rangle_{V}/\nu_t & \langle Q_{yz} \rangle_{V}/\nu_t & \langle Q_{xz} \rangle_{V}/\nu_t \\
            \noalign{\smallskip}
            \hline
	    Co1-00         &    -0.09 & \ \ 1.60 & -0.03 \\
	    Co1-00-shear   & \ \ 0.06 &    -1.37 & -0.06 \\
	    Co1-7.5        &    -4.83 & \ \ 3.10 & -4.59 \\
	    Co1-7.5-shear  &    -0.12 &    -1.17 & -0.37 \\
	    Co1-30         &    -0.86 & \ \ 7.28 & -6.64 \\
	    Co1-30-shear   & \ \ 0.07 &    -0.44 & -0.20 \\
	    Co1-60         &    -0.56 & \ \ 7.15 & -4.87 \\
	    Co1-60-shear   & \ \ 0.04 & \ \ 0.01 & -0.13 \\
	    \hline
	    Co10-00        & \ \ 0.72 &    -2.53 & \ \ 0.19 \\
	    Co10-00-shear  &    -1.39 & \ \ 7.03 &    -0.09 \\
	    Co10-7.5       & \ \ 27.4 &    -9.39 &    -1.53 \\
	    Co10-7.5-shear & \ \ 5.70 & \ \ 0.07 & \ \ 0.04 \\
	    Co10-30        & \ \ 8.68 &    -3.19 &    -8.76 \\
	    Co10-30-shear  & \ \ 1.00 & \ \ 0.05 &    -0.58 \\
	    Co10-60        & \ \ 2.62 &    -0.28 & \ \ 4.02 \\
	    Co10-60-shear  & \ \ 1.74 & \ \ 0.30 &    -0.15 \\
            \noalign{\smallskip}
            \hline
         \end{array}
     $$ 
     \vspace{-0.5cm}
   \end{table}

   \begin{figure*}
   \centering
   \includegraphics[width=0.8\textwidth]{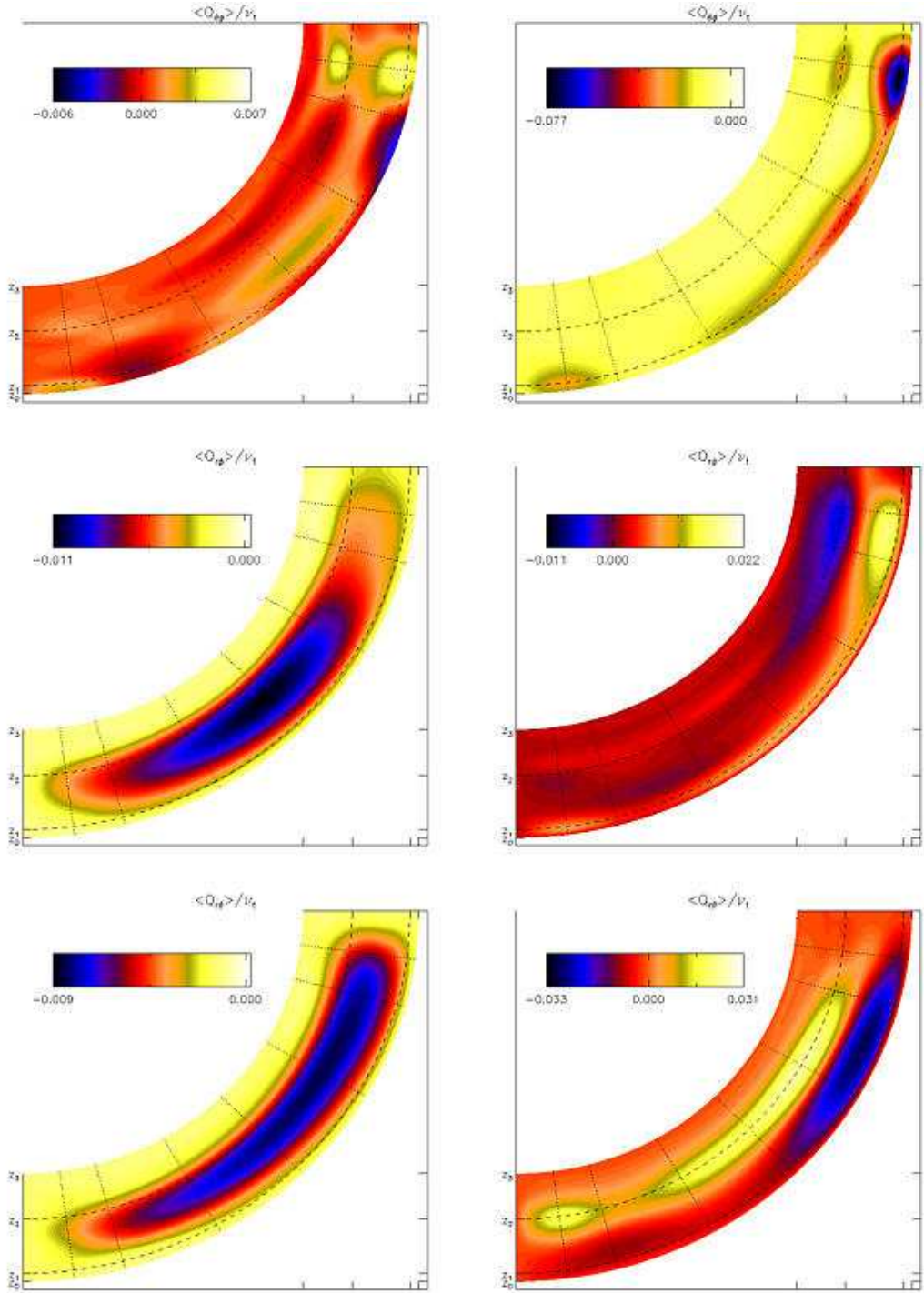}
      \caption{A polar representation of the horizontally averaged
      Reynolds stress components $\langle Q_{\theta \phi} \rangle$
      (top), $\langle Q_{r \phi} \rangle$ (middle), and $\langle Q_{r
      \theta} \rangle$ (bottom), normalised by the turbulent
      viscosity.  Data from the sets Co1 (left) and Co10 (right) is
      shown. The dashed lines denote the top and bottom of the
      convectively unstable layer, and the dotted lines indicate the
      latitudes for which data exists; to represent the latitudinal
      dependence of the stresses smoothly, a cubic spline
      interpolation between the datasets has been made. Note that the
      stresses are presented as they would appear in spherical
      coordinates on the southern hemisphere.}
      \label{pic:polar_reynolds}
   \end{figure*}

   \begin{figure}[t]
   \centering
   \includegraphics[width=0.48\textwidth]{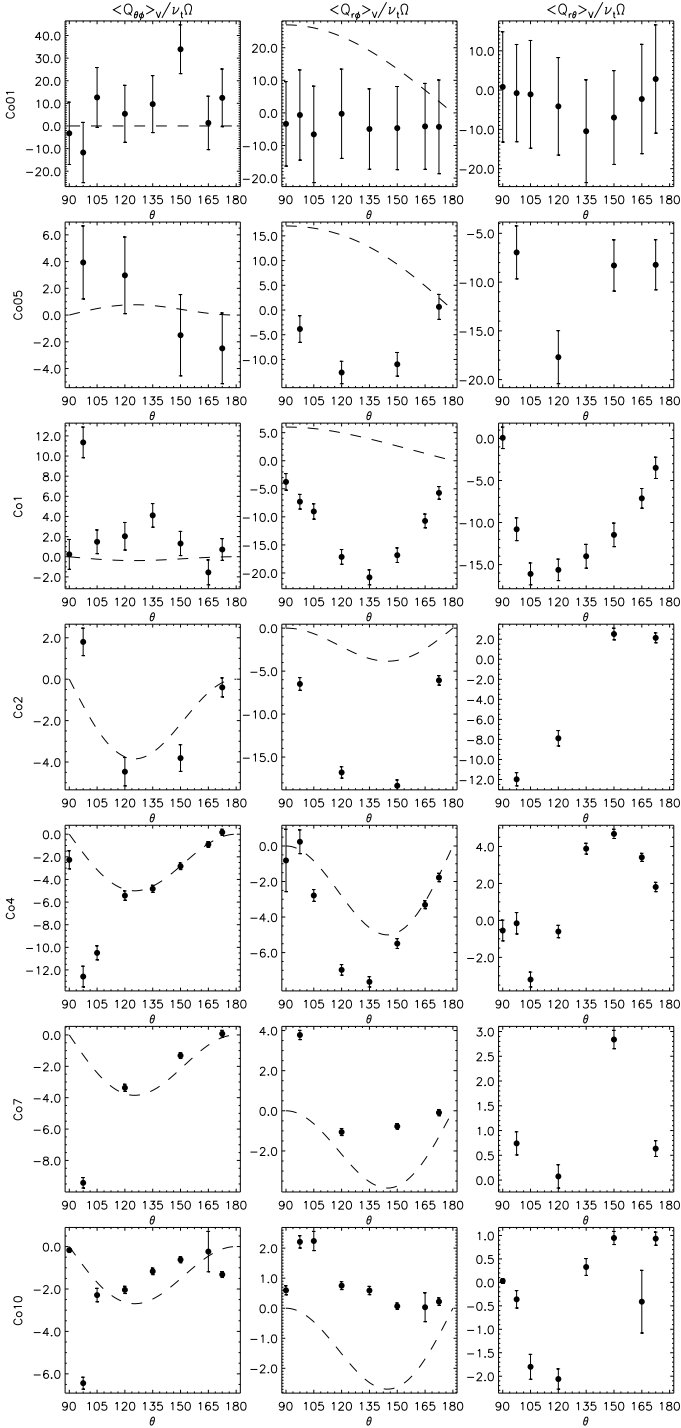}
      \caption{The volume averaged Reynolds stress components $\langle
      Q_{\theta \phi} \rangle$ (left), $\langle Q_{r \phi} \rangle$
      (middle), and $\langle Q_{r \theta} \rangle$ (right) in units of
      $[10^{-2}]$, from the Ra = $2.5 \cdot 10^{5}$ calculations,
      normalised by $\nu_{\rm t}\Omega$. The errorbars are calculated
      from Eq.~(\ref{equ:error}). For comparison, the dashed lines
      show the results of KR93 for the corresponding Coriolis number.}
      \label{pic:reynolds}
   \vspace{-0.5cm}
   \end{figure}

   \begin{figure}
   \centering
   \includegraphics[width=0.48\textwidth]{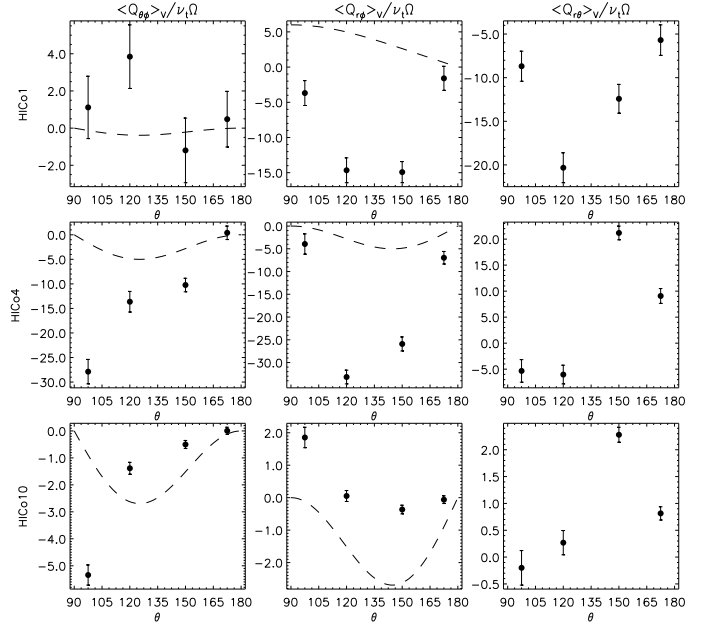}
      \caption{Same as Fig.~\ref{pic:reynolds}, but for the Ra $=
      10^{6}$ calculations.}
      \label{pic:reynoldsHI}
   \end{figure}

   \subsubsection{Radial distribution}
   We continue by discussing the vertical (radial) profiles of the
   Reynolds stresses as the Taylor number is varied. These are shown
   for the horizontally averaged off-diagonal Reynolds stress
   components, normalised by an estimate of the turbulent viscosity
   $\nu_{t} = \frac{1}{3}u_{t}d$, in Fig.~\ref{pic:rey_hori} for slow
   (set Co01), moderate (Co1), and rapid rotation (Co10). In order to
   visualise the latitudinal variation of the stresses, the data from
   the two latter sets are shown in a polar plot in
   Fig.~\ref{pic:polar_reynolds}.

   The horizontal component, $\langle Q_{xy} \rangle$, shows several
   sign changes as function of depth and no clear trends are apparent
   for slow rotation (sets Co01 to Co05). For moderate rotation (Co1
   and Co2) larger negative values tend to occur in the convection
   zone near the equator. For more rapid rotation, this latitudinal
   trend is more pronounced, and as a function of depth, a positive
   peak near the upper boundary develops. The magnitude of this
   feature increases with increasing rotation. A smaller positive peak
   develops in the overshoot region for the rapid rotation cases. The
   uppermost panels of Fig.~\ref{pic:polar_reynolds} show more clearly
   the tendency of $\langle Q_{xy} \rangle$ to obtain its largest
   value near the equator. The maxima of $\langle Q_{xy} \rangle$ also
   tends to occur in the regions where the vertical flow is deflected
   to the horizontal directions, namely in the overshoot layer
   immediately below the convection zone and near the upper boundary.
  
   The vertical Reynolds stress component, $\langle Q_{yz} \rangle$,
   shows a distinct trend of being mostly positive in the bulk of the
   convection zone for the slow and moderate rotation. Considerably
   weaker positive values of the stress occur in the overshoot region
   for these cases. The magnitude of this component increases up to
   the set Co2, after which the trend begins to change and for the set
   Co4 this quantity is more or less zero in the bulk of the
   convection zone near the equator (the positive peak in the
   overshoot layer remains) and the overall magnitude at other
   latitudes decreases. Similar trend continues in the set Co7. There
   is a sign change in the bulk (but not in the stable region) for the
   most rapid rotation (set Co10). As the latitude increases, the
   negative dip near the top of the domain becomes narrower.

   The meridional component of the stress, $\langle Q_{xz} \rangle$,
   is negative throughout the domain for the slow rotation cases Co01
   to Co1. The magnitude of this quantity is less in the overshoot
   region than in the bulk of the convection zone. For the set Co2 the
   same trend continues on the lower latitudes but there is a sign
   change at higher latitudes. With even more rapid rotation, $\langle
   Q_{xz} \rangle$ still exhibits a negative dip in the unstable
   region, but in addition a comparable positive peak appears in the
   overshoot region.

   In comparison, the depth dependencies of the Reynolds stresses are
   in many cases at variance with those obtained by Chan
   (\cite{Chan01}). The accordance seems to be the best for the slow
   rotation cases ($\rm Co \le 1$) and for the components $\langle
   Q_{yz} \rangle$ and $\langle Q_{xz} \rangle$. The sign changes
   observed when the rotation becomes large enough are seen to occur
   in the results of Chan (\cite{Chan01}) only for $\langle Q_{xy}
   \rangle$, whereas in our case all components exhibit this
   behaviour. The differences here also reflect the variance seen in
   the mean velocities.

   \begin{figure}
   \centering
   \includegraphics[width=0.5\textwidth]{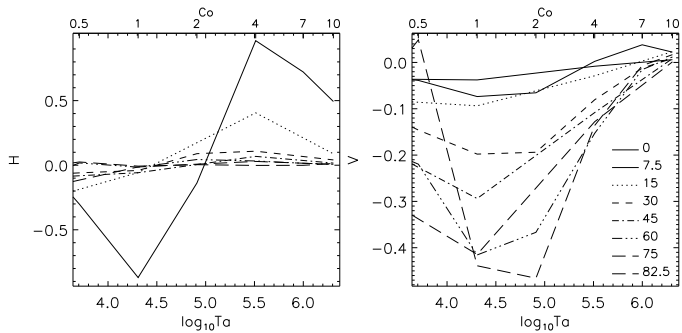}
      \caption{Coefficients $H$ (left) and $V$ (right) from different
      latitudes as function of the Taylor number.}
      \label{pic:Lquench}
   \end{figure}

   \subsubsection{Latitudinal distribution}

   Figs.~\ref{pic:reynolds} and \ref{pic:reynoldsHI} depict the volume
   averages of the Reynolds stresses normalised by $\nu_{\rm t}
   \Omega$ as functions of latitude for all sets of calculations. We
   have adopted spherical coordinates in order to facilitate
   comparisons with other studies. Here we give an error estimate
   using a modified mean error of the mean, Eq.~(\ref{equ:error}). For
   comparison, we show the results of KR93 for the corresponding
   Coriolis number for each set.

   We begin our analysis from the horizontal component $\langle
   Q_{\theta \phi} \rangle_{V}$ (shown in the left columns of
   Figs.~\ref{pic:reynolds} and \ref{pic:reynoldsHI}), which shows a
   lot of scatter and the errorbars are large for the slow rotation
   cases $\rm Co < 1$; it seems that no real effect is present for
   these cases. As the rotation is increased, a trend starts to
   appear: the sign of the stress is negative at all latitudes, and
   the maximal magnitude is obtained at the latitude of $\Theta =
   -7.5^{\degr}$ accompanied by a rapid increase near the equator and
   a slower decrease when moving towards the poles; a series expansion
   in increasing powers of $\sin^{2} \theta$ traditionally used in
   describing the latitudinal distribution (e.g. R89) fits poorly to
   these profiles (not shown). This behaviour is very similar to the
   one reported by Pulkkinen et al. (\cite{Pulkki93}) for $\rm Co
   \approx 5$ in our notation, and visible also in the results of Chan
   (\cite{Chan01}).

   For this stress component observations from the Sun exist
   (e.g. Ward \cite{Ward65}; Virtanen \cite{Virtanen89}). On the sun
   $\langle Q_{\theta \phi} \rangle_{V}$ is negative on the southern
   hemisphere with a more or less a linear variation as a function of
   latitude (Pulkkinen \& Tuominen \cite{Pulkki98}). The negative sign
   indicates that there is an angular momentum flux towards the
   equator. Qualitative correspondence with the observations is fairly
   good when $\rm Co \ge 2$.

   A comparison with KR93 shows that, for the moderate rotation cases,
   the sign of $\langle Q_{\theta \phi} \rangle_{V}$ is the same in
   both studies, but that the latitudinal distribution is much more
   concentrated towards the equator in the present results. Also the
   magnitude of the horizontal $\Lambda$-effect, $H \equiv
   \frac{\langle Q_{\theta \phi} \rangle_{V}}{\nu_{t} \cos \theta
   \Omega}$, seems to be larger in our case. We find that the absolute
   value of $\langle Q_{\theta \phi} \rangle_{V}$ increases up to the
   set Co7, but the growth seems to have saturated for the set Co10,
   where the magnitude is comparable to the Co7 set. What this implies
   is that $H$ itself is decreasing as function of $\Omega$ for the
   most rapid rotation cases. This is an indication of rotational
   quenching of the horizontal $\Lambda$-effect, see the left panel of
   Fig.~\ref{pic:Lquench}, which is noticiable when $\rm Co \ge
   4$. Recent observational studies of rapidly rotating stars seem to
   support the view that generation of horizontal differential
   rotation is strongly suppressed when rotation is large enough
   (e.g. Reiners \& Schmitt \cite{ReSch03b}).

   The stress component $\langle Q_{r \phi} \rangle_{V}$ (the middle
   columns of Figs.~\ref{pic:reynolds} and \ref{pic:reynoldsHI})
   describes the radial flux of angular momentum, and is thus vital in
   generating radial differential rotation. We find that $\langle Q_{r
   \phi} \rangle_{V}$ is mostly negative, corresponding to inward
   transport of angular momentum, for all sets except the case Co01,
   where no clear trend is seen and the two most rapid rotation cases
   where a change of sign occurs. The magnitudes at the midlatitudes
   tend to be the largest for all the sets up to Co4. As the rotation
   is increased further, the sign changes, starting from the
   equatorial regions so that for the set Co10, $\langle Q_{r \phi}
   \rangle_{V}$ is positive at all latitudes, and obtains the maximum
   near the equator. The qualitative picture of $\langle Q_{r \phi}
   \rangle_{V}$ for the sets Co1 to Co4 is in accord with Pulkkinen et
   al. (\cite{Pulkki93}) and Chan (\cite{Chan01}), whereas on the
   rapid rotation end we seem to have come to a regime not yet
   encountered in the other studies. A comparison to the results of
   KR93 seems to support this view as well. The relatively strong and
   positive vertical $\Lambda$-effect of KR93 for slow rotation is not
   present in our results. When rotation is intermediate (sets Co2 and
   Co4), the correspondence is rather good, but the sign change at
   high rotation is not seen in the study of KR93. We note here that
   absolute value of $\langle Q_{r \phi} \rangle_{V}$ does not change
   appreciably from the set Co2 to Co10, which means that the vertical
   $\Lambda$-effect $V = \frac{\langle Q_{r\phi} \rangle_{V}}{\nu_{t}
   \sin \theta \Omega}$ is also subject to strong rotational
   quenching, and that this effect is much more clear for $V$ than for
   $H$ (see the right panel of Fig.~\ref{pic:Lquench}).

   Finally, the Reynolds stress capable of driving meridional flow,
   $\langle Q_{r \theta} \rangle_{V}$, does not show as clear trends
   as the two other components. For slow rotation (sets Co01 to Co1)
   this component is negative at all latitudes. For moderate rotation
   (set Co2), the values near the pole start to get positive. Towards
   the rapid rotation end, there seems to be a negative dip at low
   latitudes ($\Theta \approx -15 \degr \ldots -30 \degr$) and a
   positive one at high latitudes ($\Theta \approx -60 \degr$). This
   qualitative trend is again similar to what Pulkkinen et
   al. (\cite{Pulkki93}) find in their study.

   Our main conclusions for the Reynolds stress components relevant for
   the generation of differential rotation indicate that the angular
   momentum is transported towards the equator when rotation is
   significant (cases Co1 and up) and radially inwards for slow and
   moderate rotation (sets Co05 to Co4), and outwards for rapid rotation
   (sets Co7 and Co10). The efficiency of the angular momentum transport
   is largest at mid-latitudes for slow and moderate rotation and near
   the equator for the rapid rotation cases. 

   The observed rotational quenching of the $\Lambda$-coefficients is
   of great interest in the context of the dynamo theory, where the
   generation of the toroidal magnetic field occurs via the
   $\Omega$-effect, dependent on the \emph{absolute} differential
   rotation $\Delta \Omega$. According to our results the generators
   of the $\Omega$-effect may be significantly reduced, the quenching
   of the $V$ coefficients being stronger and taking place with lower
   rotation rates than for the $H$ coefficients. This implies that the
   horizontal differential rotation would dominate over the radial one
   in stars with moderate rotation rates, but that even the horizontal
   differential rotation would diminish significantly in the rapid
   rotation regime. Therefore, the dynamos operating in rapidly
   rotating stars can be expected to be of $\alpha^{2}$-type. One
   must, however, note that the $\alpha$-effect is probably, at least
   at the poles, subject to rotational quenching as well (Ossendrijver
   et al. \cite{Osse01}). It remains to be seen how the reverse of the
   sign and latitude distribution of the helicity noted in
   Sect. \ref{subsec:heli} in the rapid rotation regime changes this
   picture.

   \begin{table*}
   \centering
      \caption[]{Summary of the calculations with an imposed azimuthal
      magnetic field.}
      \vspace{-0.75cm}
      \label{tab:magruns}
     $$
         \begin{array}{p{0.04\linewidth}cccccccccccc}
            \hline
            \noalign{\smallskip}
            Run      & {\rm Ch} & {\rm Rm} &H_{\rm tot} [10^{-2}] &V_{\rm tot} [10^{-1}] &M_{\rm tot} [10^{-2}] & {\rm E_{kin}}\;[10^{-3}] & {\rm E_{mag}/E_{kin}} & {u_{t}} &{b_{t}} & A_H\;[10^{-3}] & A_V\;[10^{-3}] & \Delta T/\tau\\
            \noalign{\smallskip}
            \hline
            M0       & 0             & -   &   \;\;\; 4.38 & -3.67 &  -5.80 & 2.03 & -    & 0.085  & -     &  1.6 & -12.1  &  108 \\
            M1       & 1             & 141 &         -0.01 & -3.57 &  -2.41 & 1.98 & 0.22 & 0.084  & 0.009 &  1.5 & -12.1  & 57.8 \\
            M2       & 10            & 133 &   \;\;\; 2.26 & -3.90 &  -6.77 & 1.74 & 0.50 & 0.079  & 0.021 &  2.8  &-13.0  & 54.4 \\
            M3       & 10^{2}        & 129 &         -2.94 & -5.41 & \;\;\; 8.27 & 1.51 & 1.04 & 0.073  & 0.035 &  3.5 & -10.8  & 50.5 \\
            M4       & 10^{3}        & 119 &         -9.38 & -5.90 &  -8.86 & 1.43 & 4.15 & 0.071  & 0.042 & 16.0 &  \;\;\; 7.9 & 47.7 \\
            \noalign{\smallskip}
            \hline
         \end{array}
     $$ 
     \vspace{-0.5cm}
   \end{table*}

   \begin{figure}
   \centering
   \includegraphics[width=0.5\textwidth]{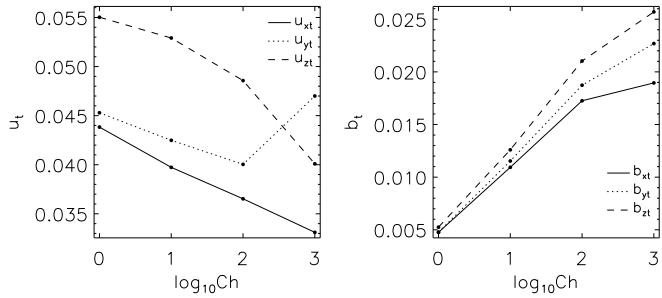}
      \caption{The volume averages of the rms-velocities (left) and
       rms-magnetic fields (right) in different directions.}
      \label{pic:aniso}
   \end{figure}

   \begin{figure}
   \centering
   \includegraphics[width=0.5\textwidth]{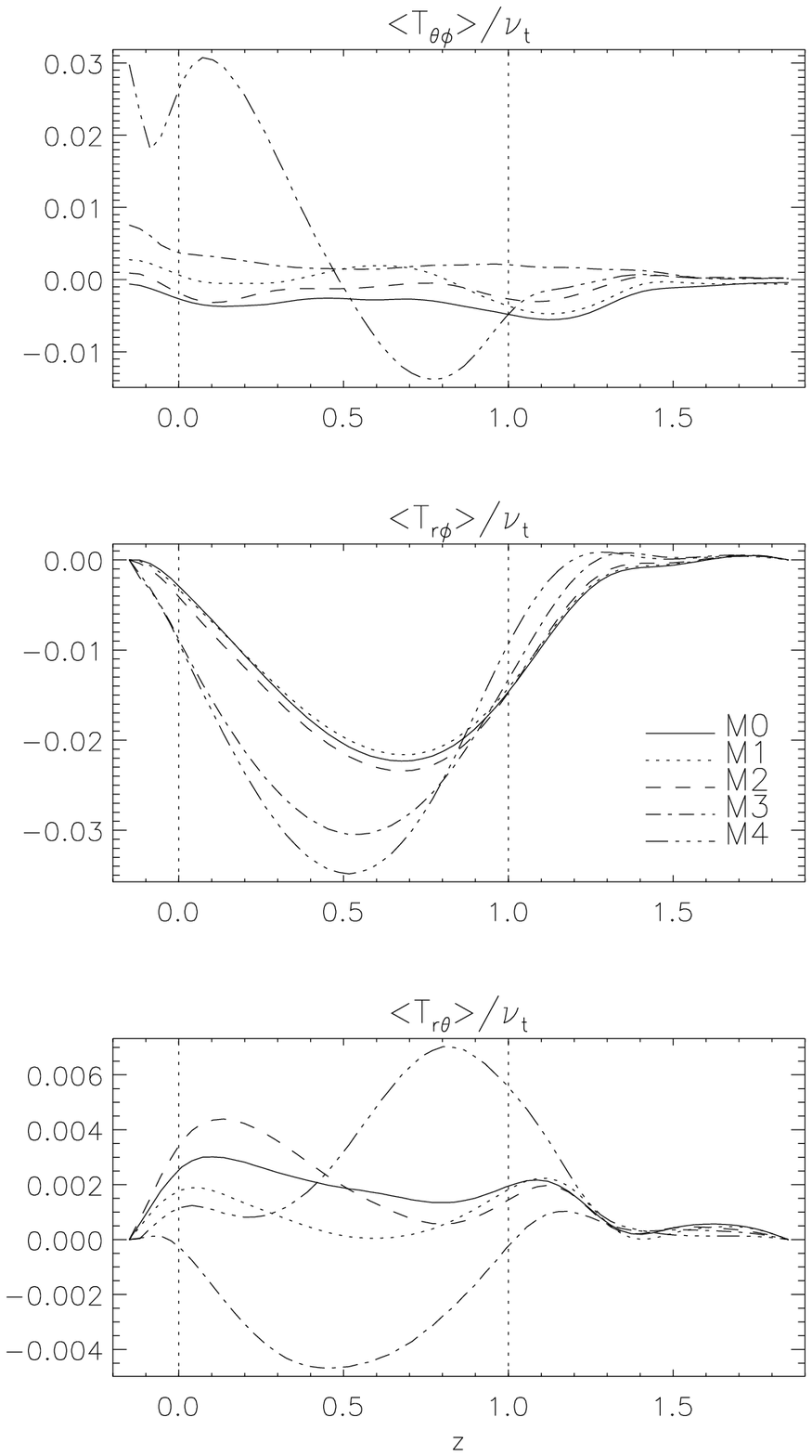}
      \caption{Vertical profiles of the horizontally averaged total
      turbulent stresses $\langle T_{ij} \rangle$ normalised by the
      turbulent viscosity from the magnetic runs M0 to M4 transformed
      into spherical coordinates.}
      \label{pic:reymag}
   \end{figure}

   \begin{figure}
   \centering
   \includegraphics[width=0.5\textwidth]{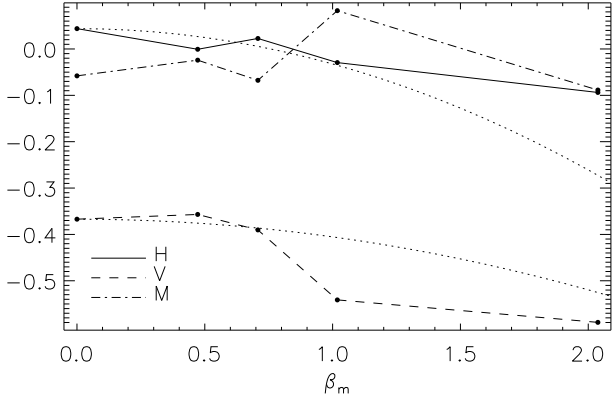}
      \caption{ $\Lambda$-coefficients $H_{\rm tot}$, $V_{\rm tot}$,
      and $M_{\rm tot}$ as functions of the normalised field strength,
      $\beta_m=|\vec{B}|/\sqrt{\mu_0 \rho u_t^2}$. The quenching functions
      derived by Kitchatinov et al. (\cite{Kitcha94b}) are overplotted
      with the dotted linesymbols (see Sect. \ref{subsubsec:maglam}
      for explanations).}
      \label{pic:VHM_mag}
   \end{figure}

   \subsubsection{The effect of an imposed azimuthal magnetic field}
   \label{subsubsec:maglam}

   We investigated the behaviour of the total turbulent
   stresses, defined as
   \begin{equation}
     T_{ij} = Q_{ij} - (\mu_0 \rho)^{-1} M_{ij},
   \end{equation}
   where $Q_{ij}$ and $M_{ij}$ are the corresponding Reynolds and
   Maxwell stress tensor components, in the magnetic regime by a set
   of calculations (listed in Table~\ref{tab:magruns}) in which we
   impose a large-scale azimuthal magnetic field ${\vec B}=(0,B_0=\rm
   const,0)$ of varying strength on top of a convection pattern at a
   saturated state. This choice seems plausible since the dominating
   component of the solar mean magnetic field is the toroidal
   one. Since the imposed field has no gradients, the large-scale
   Lorentz-force vanishes in the calculations, whilst the Maxwell
   stresses are non-zero due to the turbulent magnetic field component
   generated when the convective turbulence is acting on the mean
   field. We define the modified $\Lambda$-coefficients including the
   effect of the Maxwell stresses as
   \begin{eqnarray}
     H_{\rm tot} &=& - \frac{\langle T_{xy} \rangle_V}{\nu_t \cos{\theta} \Omega}\:, \\
     V_{\rm tot} &=& - \frac{\langle T_{yz} \rangle_V}{\nu_t \sin{\theta} \Omega}\:, \\
     M_{\rm tot} &=& \frac{\langle T_{xz} \rangle_V}{\nu_t \sin{\theta}
\cos{\theta} \Omega}\;,
   \end{eqnarray}
   and list the values obtained from the calculations in the columns
   4, 5, and 6 of Table~\ref{tab:magruns}.

   As a base calculation for this study we took the run Co2-60, where
   all the Reynolds stress components were appreciably large. The
   maximum Chandrasekhar number was chosen to be $10^{3}$ (run M4) which
   corresponds to a situation where the magnetic energy is roughly four
   times the kinetic energy of turbulence (Table~\ref{tab:magruns},
   column 8) in the saturated state. This calculation can be thought to
   represent the superequipartition case, whereas M3 corresponds to
   equipartition.

   When a magnetic field is imposed on top of the convection pattern,
   the flow undergoes a transition stage of a few tens of time units,
   after which it relaxes to another stable state; in measuring the
   quantities listed in Table~\ref{tab:magruns} we have neglected this
   transition phase. Due to the convective motions acting on the
   magnetic field, a fluctuating magnetic field is generated. The
   strength of the fluctuating field (measured by the rms-alue listed
   in column 10) grows roughly exponentially up to M3, after which the
   growth seems to become quenched. For the runs M1 to M3 the
   fluctuating field is dominating over the large-scale field,
   although the fraction of small-scale energy over the large-scale
   one is decreasing nearly exponentially when the strength of the
   imposed field is increased. For the run M4 the large-scale field is
   dominating over the fluctuating one and also tends to elongate
   convective structures along its direction. The total and
   fluctuating velocity fields, measured by the total kinetic energy
   and the rms-value listed in columns 7 and 9, respectively, become
   suppressed by the magnetic field, as expected; from M1 to M3 the
   suppression is, again nearly exponential, but seems to cut off for
   M4.

   Due to the presence of the imposed azimuthal field, an additional
   anisotropy is induced to the system. The azimuthal magnetic field
   is excerting a force opposing the movements in the directions
   perpendicular to it, which is seen as the suppression of the
   latitudinal and vertical velocity components, whilst the azimuthal
   velocity component is less affected (see Fig.~\ref{pic:aniso} panel
   on the left). For the largest field strength (run M4) the azimuthal
   velocity exceeds the other components, indicating that motions
   occur mostly along the field lines in contrast to the non-magnetic
   case (M0) in which the vertical (radial) direction was
   preferred. This anisotropy is also seen in the quantities $A_H$ and
   $A_V$ listed in columns 11 and 12, defined by Eqs.~(\ref{equ:AH})
   and (\ref{equ:AV}); the former is obtaining larger positive values
   as the magnetic field strength grows, the latter smaller negative
   values and finally changing sign when the azimuthal motions grow
   larger than the radial ones. The fluctuating magnetic field
   components (see Fig.~\ref{pic:aniso} panel on the right) generated
   by the convective motions rather closely reflect this anisotropy
   bar the run M4, so that, e.g., the velocity component containing
   the most of energy generates the strongest fluctuating magnetic
   field in its direction.

   The generated total turbulent stresses $T_{ij}$, plotted in
   Fig.~\ref{pic:reymag}, show rather an unexpected
   behaviour. Although, in general, the fluctuating velocity field
   becomes suppressed, the turbulent correlations are rather enhanced
   in magnitude when the strength of the imposed field is
   increased. In the hydrodynamic case M0, the stress components
   responsible for the generation of differential rotation, $\langle
   T_{r \phi} \rangle$ and $\langle T_{\theta \phi} \rangle$, are
   negative throughout the domain, the former being the strongest in
   magnitude. For this component the profile retains its shape in the
   magnetic regime, only the magnitude increasing as magnetic field
   grows. The growth is accompanied by the increase of the mean
   latitudinal flow $\langle u_x \rangle$, so that it is roughly twice
   as strong in the case M4 than in M0. The two weak negative peaks
   seen in $T_{\theta \phi}$ develop in a less systematic
   manner, the one at the interface of the convective and overshoot
   layers remains more or less unaffected by the growing magnetic
   field, but the one near the surface changes its sign and gains in
   magnitude. The behaviour of $T_{r \theta}$ does not show
   any clear trend, but changes its magnitude, profile and sign rather
   irregularly. The mean azimuthal flow, generated by the $r \theta$
   component, also lacks a systematic behaviour, although it tends to
   change its sign averaged over depth, being preferably positive in
   the regime of strong magnetic field.

   The coefficients $H_{\rm tot}$, $V_{\rm tot}$, and $M_{\rm tot}$,
   calculated from the horizontal averages of the total stresses, are
   plotted in Fig.~\ref{pic:VHM_mag} as functions of the normalised
   field strength $\beta_m = |\vec{B}|/\sqrt{\mu_0 \rho u_t^2}$, where
   $|\vec{B}| = B_0 + b_{t}$. In the hydrodynamic case M0 (see
   Fig.\ref{pic:reynolds} fourth row) the horizontal
   $\Lambda$-coefficient is positive, whilst the vertical coefficient
   and the quantity $M$ are negative. As the magnetic field is applied
   to the system, the horizontal coefficient changes its sign and
   increases monotonically in magnitude (the dashed line in
   Fig.~\ref{pic:VHM_mag}), the growth being somewhat reduced in the
   run M4. The vertical coefficient (the solid line) retains its
   negative sign, and also shows a monotonic increase in magnitude,
   with less evident signs of a suppression of the growth in the
   strong field limit. No obvious trend is visible for the quantity
   $M$ (dashed-dotted line), as already implied by the irregular
   behaviour of the stress $\langle T_{r \theta} \rangle$.

   \begin{figure}
   \centering
   \includegraphics[width=0.5\textwidth]{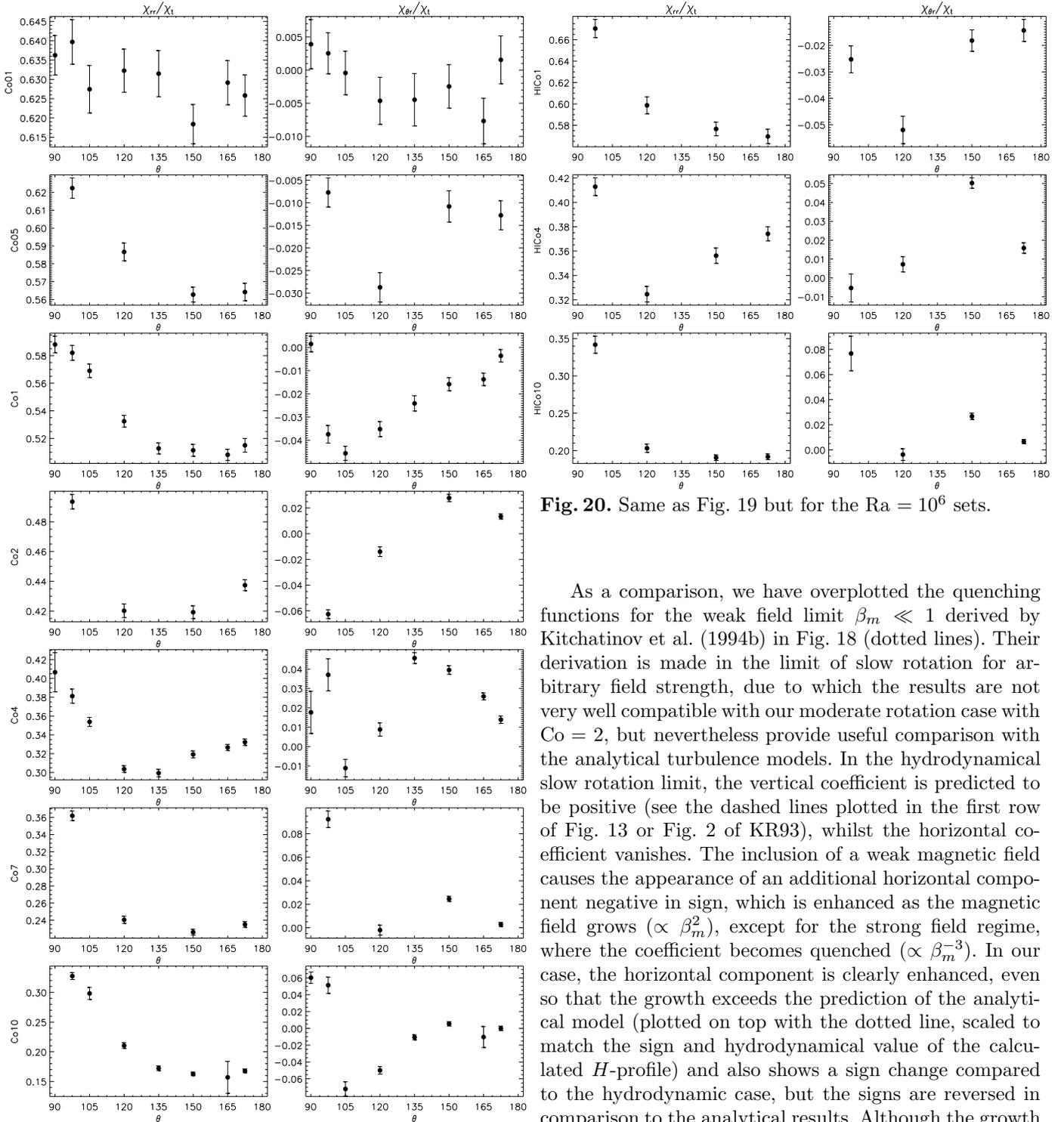}
      \caption{Volume averaged eddy heat flux tensor components
      $\chi_{rr}/\chi_{t}$ (left), and $\chi_{\theta r}/\chi_{t}$
      (right) as functions of latitude for the Ra $= 2.5 \cdot 10^{5}$
      runs. The errorbars denote the modified mean error
      of the mean as calculated from Eq.~(\ref{equ:error}).}
       \label{pic:tuhe}
   \end{figure}

   \begin{figure}
   \centering
   \includegraphics[width=0.5\textwidth]{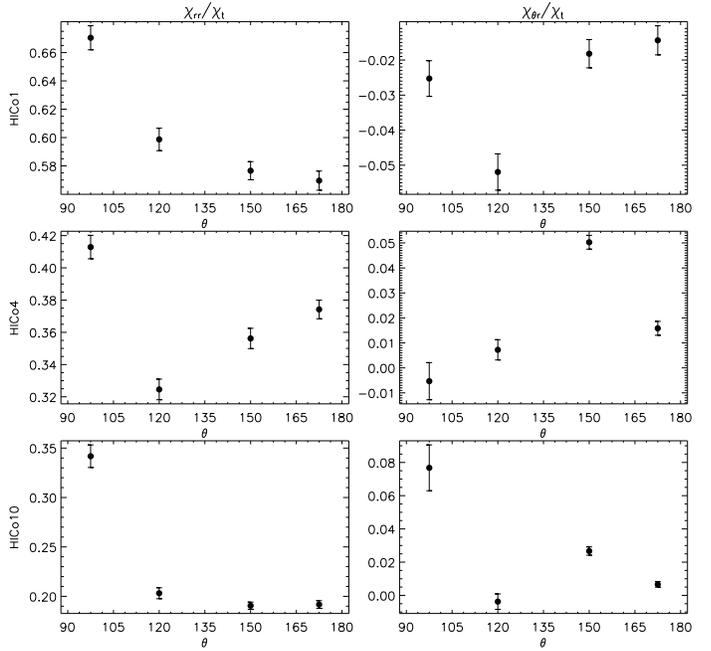}
      \caption{Same as Fig.~\ref{pic:tuhe} but for the Ra $= 10^{6}$
      sets.}
       \label{pic:tuheHI}
   \end{figure}

   As a comparison, we have overplotted the quenching functions for
   the weak field limit $\beta_m \ll 1$ derived by Kitchatinov et
   al. (\cite{Kitcha94b}) in Fig.~\ref{pic:VHM_mag} (dotted
   lines). Their derivation is made in the limit of slow rotation for
   arbitrary field strength, due to which the results are not very
   well compatible with our moderate rotation case with $\rm Co = 2$,
   but nevertheless provide useful comparison with the analytical
   turbulence models. In the hydrodynamical slow rotation limit, the
   vertical coefficient is predicted to be positive (see the dashed
   lines plotted in the first row of Fig.~\ref{pic:reynolds} or Fig. 2
   of KR93), whilst the horizontal coefficient vanishes. The inclusion
   of a weak magnetic field causes the appearance of an additional
   horizontal component negative in sign, which is enhanced as the
   magnetic field grows $(\propto \beta_{m}^2)$, except for the strong
   field regime, where the coefficient becomes quenched $(\propto
   \beta_{m}^{-3})$. In our case, the horizontal component is clearly
   enhanced, even so that the growth exceeds the prediction of the
   analytical model (plotted on top with the dotted line, scaled to
   match the sign and hydrodynamical value of the calculated
   $H$-profile) and also shows a sign change compared to the
   hydrodynamic case, but the signs are reversed in comparison to the
   analytical results. Although the growth becomes somewhat suppressed
   in the strong field limit, it is significantly milder than
   predicted by the analytical investigation. A quenching function of
   a similar form (plotted on top of the calculated $V$-profile with a
   dotted line, scaled to match the hydrodynamical value) was derived
   for the vertical component, accompanied by a sign change from
   positive to negative with increasing magnetic field. In our
   calculations the vertical component remains negative and increases
   in magnitude as function of $\beta_m$. The growth is also much
   stronger than the one predicted by the analytical expression and no
   significant quenching, as expected from the analytical theory
   ($\propto \beta_{m}^{-3}$), however, is found in the strong field
   regime.

   In summary, the produced set of magnetic runs with an imposed
   azimuthal field indicate that the $\theta \phi$ and $r \phi$
   components of the turbulent stresses, capable of generating
   horizontal and vertical differential rotation, respectively, increase
   in magnitude with increasing imposed field strength whilst the $r
   \theta$ component, capable of affecting the meridional circulation
   pattern, shows no clear trends although also gains its maximum value
   for the case of the strongest magnetic field. The model setup,
   however, hardly represents the situation in a real stellar convection
   zone, where magnetic fields arise due to the turbulent convective and
   large-scale shearing motions, exhibiting gradients and thereby mean
   electric currents leading to the Malkus-Proctor effect, which has
   been excluded from this investigation.

   \subsection{Turbulent heat transport}

   In Figs.~\ref{pic:tuhe} and ~\ref{pic:tuheHI} we show the latitudinal
   variation of the volume averaged eddy heat flux tensor components
   $\chi_{rr}$ and $\chi_{\theta r}$, defined via
   Eq.~(\ref{equ:deftuhe}). We use an estimate of the turbulent
   conductivity $\chi_{t} = \nu_{t}$ as normalisation. The error is
   estimated as in Figs.~\ref{pic:reynolds} and \ref{pic:reynoldsHI}.

   The radial component $\chi_{rr}$ is always positive, indicating that
   the heat flux is directed radially outwards, except at the equator
   for the case of highest rotation (Co10-00). In this calculation the
   temperature stratification becomes distorted by the strong rotational
   influence so that $\delta_r$ becomes negative. The latitudinal trends
   for $\chi_{rr}$ are clear: maximum occurs at the equator with
   diminishing magnitude towards the pole and with increasing rotation
   rate. There seems to be, in general, a tendency for the minimum to
   occur at a latitude $\Theta \approx -45\degr \ldots -60 \degr$,
   indicating a minimum in the efficiency of convection. These trends
   concur with the 'deeper layer' results of Pulkkinen et
   al. (\cite{Pulkki93}). However, the occurence of the minimum in the
   turbulent heat flux at mid-latitudes seems to contradict the one
   obtained from the Nusselt numbers (see Fig.~\ref{pic:nusselt}), which
   indicates that the efficiency of convection decreases monotonically
   from the equator towards the pole. This discrepancy can be explained
   as follows: the value of the Nusselt number, Eq.~(\ref{equ:nusselt}),
   depends on only $F_{\rm rad}^{(0)} = \kappa \frac{\Delta \langle e
   \rangle}{\Delta z}$, where the difference $\Delta$ is taken using the
   values at the layers $z_1$ and $z_2$. In other words, the temperature
   stratification is assumed to be linear. If this was true, the radial
   superadiabacity, $\delta_r$, would be proportional to $F_{\rm
   rad}^{(0)}$, and the latitudinal profiles of $\chi_{rr}$ would show
   the same qualitative trend as the Nusselt number since $\langle
   u_{z}' T' \rangle$ itself and Nu show similar tendencies as function
   of latitude. The superadiabatic layer, however, gets shallower from
   the equator to the pole, which affects $F_{\rm rad}^{(0)}$
   immediately because it depends only on the values of $e$ at depths
   $z_1$ and $z_2$. On the other hand, the convection zone is more
   superadiabatic (i.e $\delta_r$ is larger) at mid-latitudes than near
   the poles. Since $\chi_{rr}$ takes into account the whole
   stratification rather than just the endpoints, it probably gives a
   more reliable estimate of the convection efficiency than the Nusselt
   number does.

   The $\chi_{\theta r}$ component shows a less coherent picture. For
   slow and moderate rotation $\chi_{\theta r}$ is negative throughout
   the latitude range, the maximum being at low latitudes ($\Theta
   \approx -10 \ldots -30 \degr$). For the case Co4 $\chi_{\theta r}$
   obtains positive values near the pole and for the higher rotation
   cases there is a definite positive peak at high latitudes, and
   another near the equator.

   Due to the boundary conditions we cannot actually have a temperature
   difference between the top layers of the boxes situated at different
   latitudes, but the radial gradient and the value at the bottom of the
   box are not constrained. In accordance with the latitudinal
   distribution of $\chi_{rr}$, we find that the temperature gradient in
   the equatorial cases is always shallower than at higher latitudes,
   and that the mid-latitudes do seem to be the least favoured in the
   sense that there the gradient is the steepest. If this situation was
   naively moved onto a spherical star where we assume the core to have
   a constant surface temperature below the convection zone, this would
   result in a warm equator and pole, and a cooler region at
   mid-latitudes. However, this approach is probably too simple since it
   does not take into account the latitudinal transport which shows a
   more complex pattern. For slow rotation (Co01 to Co1), the negative
   values of $\chi_{r \theta}$ indicate an equatorward heat flux at all
   latitudes. In the intermediate regime (Co2 and Co4) there is a
   poleward flux at the polar regions and a less pronounced equatorward
   one near the equator, supporting the view of the cooler mid-latitudes
   discussed above. This tendency, however, is reversed in the rapid
   rotation regime, so that a poleward flux appears at the equatorial
   regions, and an equatorial one at high latitudes, which kind of a
   flux would act as smoothing out the latitudinal temperature
   difference generated by the radial heat flux.


\section{Conclusions}
\label{sec:conclu}
   We have performed local three-dimensional hydrodynamical
   calculations of compressible convection with the inclusion of an
   overshoot layer in Cartesian volumes placed at various latitudes
   and with largely varying rotation. We find that the convective
   structure changes from an ordered cellular pattern of slow and
   moderate rotation (Co $\le 2$) to a highly irregular small-scale
   pattern for rapid rotation (Co $\ge 4$). The efficiency of
   convection is noted to, in general, decrease as function of
   increasing rotation, while the latitudinal dependence indicates a
   minimum at mid-latitudes and more efficent convection at the
   equator and near the poles.

   The kinetic helicity, $\mathcal{H}$, shows a similar trend as
   previous numerical studies of magnetoconvection (e.g. Ossendrijver
   et al. \cite{Osse02}) and helioseismic inversions (Duvall \& Gizon
   \cite{DuGi00}) or large positive values at the south pole and an
   approximately $\cos \theta$-latitude dependence for slow and
   moderate rotation. We, however, find that for rapid enough (Co $\ge
   4$) rotation the sign of the helicity changes and the latitudinal
   distribution follows more a $\sin \theta$-dependence at latitudes
   $\Theta \le -15\degr$ than the $\cos \theta$-profile typical for
   the two aforementioned studies. It has been concluded that the
   $\alpha$ computed from magnetoconvection calculations qualitatively
   agrees with the first order smoothing result $\alpha \approx
   -\frac{1}{3} \tau' \mathcal{H}$ (e.g. Ossendrijver et
   al. \cite{Osse01}). If this behaviour carries over to the rapid
   rotation regime, the $\alpha$-effect would also be subject to a
   sign change and simultaneous decrease of magnitude at higher
   latitudes.

   The convective overshooting is reduced monotonically as a function
   of rotation. The latitudinal dependence shows two regimes; (i) for
   slow and intermediate rotation the overshooting depth decreases as
   latitude increases, and that (ii) for rapid rotation the trend is
   the opposite. Imposing an azimuthal magnetic field causes the
   overshooting depth to decrease monotonically as a function of
   increasing field strength.

   The mean flows, generated by the Reynolds stresses, also show two
   distinct regimes; in cases of slow and intermediate rotation the
   latitudinal flow is poleward and the azimuthal one retrograde. Both
   components exhibit a more or less linear shear profile in the
   convectively unstable region. For rapid rotation, the signs of the
   mean flows change so that there is a strong equatorward flow in a
   narrow region near the surface and a weak poleward flow in the bulk
   of the convection zone. The azimuthal flow is prograde in the bulk
   and with a retrograde dip near the upper boundary.
 
   The volume-averaged Reynolds stress components, $\langle Q_{\theta
   \phi} \rangle_{V}$ and $\langle Q_{r \phi} \rangle_{V}$,
   responsible for the generation of horizontal and vertical
   differential rotation, respectively, indicate that the angular
   momentum is transported equatorwards in all cases where rotation is
   significant (a positive horizontal $\Lambda$-effect). The radial
   transport occurs inwards (negative vertical $\Lambda$-effect) for
   slow and moderate rotation and towards the surface for rapid
   rotation (positive vertical $\Lambda$-effect). The latitudinal
   distribution shows that the inward transport in situated mostly at
   high latitudes and the outward transport near the equator. The
   meridional Reynolds stress component, $\langle Q_{r \theta}
   \rangle_{V}$ is negative for slow rotation and reaches positive
   values at high latitudes near the pole. We find that both
   $\Lambda$-effects are subject to rotational quenching, indicating
   the suppression of differential rotation in rapidly rotating
   stars. Comparison to the analytical turbulence model of KR93, based
   on the quasi-linear approach, shows a rather close agreement for
   the horizontal $\Lambda$-effect near the polar regions, whilst the
   values obtained from our calculations are systematically larger
   near the equator and less suppressed as the rotation is
   increased. The regime of positive vertical $\Lambda$-effect for
   slow rotation is not found from our calculations, but for
   intermediate rotation the agreement with KR93 is fairly good. For
   the rapid rotation regime, however, the positive vertical
   $\Lambda$-effect obtained in this study is in disagreement with the
   analytical theory.

   We have also produced a set of magnetohydrodynamical calculations by
   imposing a mean azimuthal magnetic field on a saturated convection
   pattern with $\rm Co=2$. As the imposed field does not have any
   gradients, no mean currents are present in the system, due to which
   the large-scale Lorentz force vanishes. The action of the convective
   motions on the mean azimuthal field, however, leads to the generation
   of a fluctuating component of the magnetic field and non-zero Maxwell
   stresses. In the magnetohydrodynamical regime both the Reynolds and
   Maxwell stresses contribute to the angular momentum transport; the
   magnitude of the total turbulent stresses is observed to increase
   monotonically, although the turbulent velocity field, in general,
   becomes suppressed. The horizontal $\Lambda$-effect is observed to
   change sign from positive to negative as the field strength
   increases, indicating a new regime of poleward transport due to the
   presence of the magnetic field. 

   The turbulent heat transport indicates a rather strong latitudinal
   dependence of the radial heat flux, whereas the latitudinal flux
   shows no distinct tendencies. A simple analysis of the radial heat
   flux indicates that the equatorial regions should be the warmest,
   followed by the polar regions, and a cool belt at mid-latitudes
   would appear. However, this interpretation is most probably based
   on too simple assumptions since the thermal structure is also
   affected by the latitudinal heat flux and the meridional
   circulation. On the other hand, the latter depends on the thermal
   structure so a straightforward analysis as above is inadequate. 

   Having the knowledge of the Reynolds stresses and turbulent heat
   transport as functions of rotation and latitude it is, in
   principle, possible to investigate the implications of these local
   results on the global dynamics (rotation law, thermal structure) by
   utilizing a mean-field model in a spherical geometry. However, this
   analysis is not within the scope of this paper.

\begin{acknowledgements}
      The calculations were carried out using the supercomputers
      hosted by \emph{CSC-Scientific Computing Ltd.}, Espoo,
      Finland. PJK acknowledges the financial support from Magnus
      Ehrnrooth Foundation and the Finnish graduate school for
      Astronomy and Space Physics, and the travel assistance from the
      DFG graduate school 'Nonlinear Differential Equations:
      Modelling, Theory, Numerics, Visualisation'. MJK acknowledges
      financial support from the Academy of Finland for the project
      'Modelling and interpretation of MHD phenomena in astrophysical
      flows', and from the Wihuri and V\"ais\"al\"a foundations. MJK
      is indebted for the hospitality of LAOMP Toulouse and the
      Kiepenheuer Institut f\"ur Sonnenphysik for their hospitality
      during her visits. The authors thank Michael Stix, Mathieu
      Ossendrijver, Wolfgang Dobler, Jaan Pelt, and Michel Rieutord
      from fruitful discussions and helpful comments on the
      manuscript. IT acknowledges G\"unther R\"udiger for introducing
      him to the secrets of the $\Lambda$-effect in a Babelsberg
      Bierstube two decades ago. The authors thank the referee (Axel
      Brandenburg) for the useful comments on the manuscript.
\end{acknowledgements}

\end{document}